\documentclass[longauth]{aa}

\usepackage{graphicx}
\usepackage{txfonts}
\usepackage[dvipsnames]{xcolor}
\usepackage{natbib}
\usepackage{longtable}
\usepackage{siunitx}
\usepackage{multirow}
\usepackage{hyperref}
\usepackage[normalem]{ulem}
\usepackage{mathrsfs}
\usepackage{lineno}

\defcitealias{LopezSanJuan2015}{LS15}

\hypersetup{
  colorlinks= true,   
  urlcolor  = blue,     
  linkcolor = blue, 
  citecolor = blue
}

\makeatletter
\renewcommand*\aa@pageof{, page \thepage{} of \pageref*{LastPage}}
\makeatother

\bibpunct{(}{)}{;}{a}{}{,}    

\usepackage[english]{babel}
\usepackage{orcidlink}

\begin{document}

\title{J-PAS: Exploring the connection between local environment and the triggering of bursty star formation in Extreme Emission Line Galaxies}
\titlerunning{Environment of EELGs in J-PAS}

\author{Mauro González-Otero \inst{1}\thanks{email: \texttt{mgonzalez@iaa.csic.es} and \texttt{mauromarago@gmail.com}}\orcidlink{0000-0002-4837-1615}
        \and
        Jose M. Vílchez \inst{1}\orcidlink{0000-0001-7299-8373}
        \and
        Antonio Giménez-Alcázar  \inst{1}\orcidlink{0000-0001-5891-1083}
        \and
        Carlos López-Sanjuan \inst{2,3}\orcidlink{0000-0002-5743-3160}
        \and
        Ricardo Amorín \inst{1}\orcidlink{0000-0001-5758-1000}
        \and
        Antonio Arroyo-Polonio \inst{4}\orcidlink{0000-0002-9523-8016}
        \and
        Juan A. Fernández-Ontíveros \inst{2,3}\orcidlink{0000-0001-9490-899X}
        \and
        Sabyasachi Goswami \inst{1}
        \and
        Antonio Hernán-Caballero \inst{2,3}\orcidlink{0000-0002-4237-5500}
        \and
        Jorge Iglesias-Páramo \inst{1}\orcidlink{0000-0003-2726-6370}
        \and
        Yolanda Jiménez-Teja \inst{1}\orcidlink{0000-0002-6090-2853}
        \and
        Carolina Kehrig \inst{1,5}\orcidlink{0000-0003-1231-1482}
        \and
        Raúl González-Díaz \inst{1}
        \and
        Enrique Pérez-Montero \inst{1}\orcidlink{0000-0003-3985-4882}
        \and
        Alvaro Alvarez-Candal \inst{1}
        \and
        Iris Breda \inst{6}\orcidlink{0000-0002-3764-5780}
        \and
        Ana L. Chies-Santos \inst{7}\orcidlink{0000-0003-3220-0165}
        \and 
        Luis A. Díaz-García \inst{1}\orcidlink{0000-0001-7639-9132}
        \and
        Rosa M. González-Delgado \inst{1}\orcidlink{0000-0003-1058-1577}
        \and
        Ginés Martínez-Solaeche\inst{1}\orcidlink{0000-0002-8812-1135}
        \and
       Isabel Márquez \inst{1}\orcidlink{0000-0003-2629-1945}
       \and
       Francisco Javier Sáez Ruiz \inst{2}
        \and
       David Sobral \inst{8,9}\orcidlink{0000-0001-8823-4845}
       \and
       Eduardo Telles \inst{5}\orcidlink{0000-0002-8280-4445}
       \and 
       Stefano Zarattini \inst{2}\orcidlink{0000-0002-4877-8749}
       \and
       Zhen-Ya Zheng \inst{10}\orcidlink{0000-0002-9634-2923}
       \and
       Alejandro Lumbreras-Calle \inst{2,3}
       \and
       Raul Abramo \inst{11}\orcidlink{0000-0001-8295-7022}
       \and 
       Jailson Alcaniz \inst{5} \orcidlink{0000-0003-2441-1413}
       \and 
       Narciso Benitez \inst{1}\orcidlink{0000-0002-0403-7455}
       \and 
       Silvia Bonoli \inst{12,2}\orcidlink{0000-0002-6381-2052}
       \and 
       Saulo Carneiro \inst{5}\orcidlink{0000-0001-7098-383X}
       \and 
       Javier Cenarro \inst{2,3}\orcidlink{0000-0002-2573-2342}
       \and 
       David Cristóbal-Hornillos \inst{2}
       \and 
       Simone Daflon \inst{5}\orcidlink{0000-0001-9205-2307}
       \and 
       Renato Dupke \inst{5}
       \and 
       Alessandro Ederoclite  \inst{2}\orcidlink{0000-0003-3656-5524}
       \and 
       Carlos Hernández–Monteagudo \inst{13,14}\orcidlink{0000-0001-5471-9166}
       \and 
       Jifeng Liu \inst{15}
       \and 
       Antonio Marín-Franch \inst{2,3}\orcidlink{0000-0002-9026-3933}
       \and 
       Claudia Mendes de Oliveira \inst{11}\orcidlink{0000-0002-5267-9065}
       \and 
       Mariano Moles \inst{2}\orcidlink{0009-0002-6052-8723}
       \and 
       Fernando Roig \inst{5}\orcidlink{0000-0001-7059-5116}
       \and 
       Laerte Sodré Jr. \inst{11}\orcidlink{0000-0002-3876-268X}
       \and 
       Keith Taylor \inst{16}
       \and 
       Jesús Varela \inst{2}\orcidlink{0000-0003-0286-5940}
       \and 
       Héctor Vázquez Ramió \inst{2,3}\orcidlink{0000-0003-3135-2191}
       \and 
       Javier Zaragoza-Cardiel \inst{2,3}\orcidlink{000-0001-8216-9800}
      }

\institute{Instituto de Astrofísica de Andalucía-CSIC, 
           Glorieta de la Astronomía, E-18008 Granada, 
           Spain 
    \and Centro de Estudios de Física del Cosmos de Aragón, 
         Plaza de San Juan, 1, 44001 Teruel, 
         Spain 
    \and Unidad Asociada CEFCA-IAA, CEFCA, 
         Unidad Asociada al CSIC por el IAA y el IFCA, 
         Plaza San Juan 1, 44001 Teruel, 
         Spain 
    \and INAF–Osservatorio Astronomico di Roma,
         via di Frascati 33, I-00078 Monte Porzio Catone, 
         Italy 
    \and Observat\'orio Nacional/MCTIC, 
         R. Gen. Jos\'e Cristino, 77, 20921-400, 
         Rio de Janeiro, Brazil 
    \and Instituto de Astrof\'{i}sica e Ci\^{e}ncias do Espaço - Centro de Astrof\'isica da Universidade do Porto, 
        Rua das Estrelas, 4150-762 
        Porto, Portugal 
    \and Departamento de Astronomia, Instituto de Física, Universidade Federal do Rio Grande do Sul (UFRGS), 
         Porto Alegre, 
         R.S 90040-060, Brazil 
    \and Departamento de Física, Faculdade de Ciências, Universidade de Lisboa, 
          Edifício C8, Campo Grande, 1749-016 
          Lisbon, Portugal 
    \and BNP Paribas Corporate \& Institutional Banking, 
          Torre Ocidente Rua Galileu Galilei, 1500-392 
          Lisbon, Portugal 
    \and CAS Key Laboratory for Research in Galaxies and Cosmology, 
         Shanghai Astronomical Observatory, 
         Shanghai 200030, China 
    \and Instituto de Física, Universidade de São Paulo (IF/USP),
          São Paulo, Brazil 
    \and Donostia International Physics Center (DIPC),
          Manuel Lardizabal Ibilbidea, 4,
          San Sebastián, Spain 
    \and Instituto de Astrofísica de Canarias, 
           E-38205 La Laguna, 
           Tenerife, Spain 
    \and Departamento de Astrofísica, Universidad de La Laguna (ULL), 
            E-38205 La Laguna, Tenerife, 
           Spain 
    \and National Astronomical Observatory of China,  Chinese Academy of Sciences,
           Beijing, China 
    \and Instruments4, 4121 Pembury Place, La Canada Flintridge,
           CA 91011, U.S.A. 
\\
}        
\date{Received ---; accepted ---}
    
 
  \abstract
   {Extreme Emission Line Galaxies (EELGs) are typically compact, low-mass systems undergoing intense bursts of star formation, whose spectra are dominated by very strong nebular emission lines. Their physical conditions resemble those of galaxies in the early Universe, making them ideal analogues for investigating the processes that regulate star formation, feedback, and chemical enrichment at low redshift.}
   {Our objective is to characterise the local environment of a sample of EELGs identified in the J-PAS survey in order to determine whether their extreme properties 
   are associated with the presence of nearby companions, as expected if galaxy–galaxy interactions were the main trigger of the burst.}
   {Using the J-PAS internal data release, we analyse a sample of 917 EELGs in the redshift range of $0.03\lesssim z\lesssim0.9$, previously selected using photometric and machine-learning methods over an area of $\sim$30 deg$^2$. We employed the Pair Probability Function formalism, which combines the full photometric redshift probability distributions of galaxy pairs, to statistically identify companions within projected separations up to $200\,\mathrm{kpc}$ and velocity differences up to $500\,\mathrm{km\,s^{-1}}$. Stellar masses, star formation rates, dust attenuation, and metallicity estimates were derived for the EELGs and their companions through spectral energy distribution fitting using \texttt{CIGALE}, and Monte Carlo simulations were performed to correct for projection and completeness effects.}
   {The derived average pair fraction of the EELGs is $f_\mathrm{pair} = 0.162^{+0.048}_{-0.033}$,
   indicating that only $\sim$16\% of the sample has detectable companions, a value lower than the observed average for other star-forming galaxies within the adopted limits. When companions are identified within the sensitivity
   and stellar-mass completeness
   limits of J-PAS, they tend to be more massive and 
   to show higher SED-based metallicity estimates
   than the EELGs, suggesting that the latter are typically low-mass components in asymmetric pairs. 
   However, lower-mass companions below these limits may remain undetected.
   Finally, no statistically significant correlation is found between the equivalent widths of the [\ion{O}{III}]$\,\lambda5007$, H$\beta$, and H$\alpha$ emission lines and the projected separation to the detected companions within the limits of the J-PAS data.}
  {}


\maketitle
\nolinenumbers

\section{Introduction} \label{sec:1}
Extreme Emission Line Galaxies (EELGs) are compact, low-mass systems, usually below $10^9\,\mathrm{M}_\odot$, characterised by exceptionally large rest-frame equivalent widths (EWs) of nebular emission lines such as [\ion{O}{III}]$\,\lambda5007$, H$\beta$, and H$\alpha$, typically reaching hundreds of angstroms. This definition naturally selects galaxies that undergo extremely recent and intense bursts of star formation, displaying compact sizes ($R_{\mathrm{eff}} \lesssim 1$~kpc), specific star formation rates (sSFRs) up to two orders of magnitude above the star-forming main sequence, gas-phase metallicities well below solar, high ionisation parameters, and hard radiation fields \citep[e.g.][]{Cardamone2009, Amorin2012, Jaskot2013, Amorin2014, Amorin2015, Tang2019, Mingozzi2022, Boyett2024, Breda2024, Llerena2024}. 

In the nearby Universe, EELGs have been identified under a variety of observational labels, including blue compact dwarf (BCD) galaxies, compact \ion{H}{II} galaxies, Green Pea galaxies (GPs), and more extreme systems such as Blueberry galaxies \citep[e.g.][]{Sargent1970, Thuan1981, Terlevich1991, Kunth2000, Cardamone2009, Yang2017}. Despite the different discovery methods, redshift ranges, and selection criteria, these classes are widely interpreted as manifestations of the same underlying phenomenon: low-mass galaxies observed during a brief, extreme star-forming phase.

EELGs are currently recognised as a key galaxy population for studying galaxy evolution across cosmic time. At high redshift, galaxies with extreme nebular emission are frequently identified in deep JWST surveys and are thought to play a central role in the early assembly of galaxies, the accumulation of stellar mass, and the chemical enrichment of the intergalactic medium \citep[e.g.][]{Schaerer2022, Rhoadds2023, DelMoral2024}. In this sense, local EELGs provide valuable empirical constraints on cosmological models of star formation and feedback in dwarf galaxies \citep[e.g.][]{LLerena2023,Amorin2024,Saldadna2025,Leon2025}, which are predicted to dominate the ionising photon budget at high redshift \citep[e.g.][]{Madau2014, Robertson2015, Finkelstein2019, Matthee2022,Nadiu2022}. Moreover, the study of their gas-phase abundances and ionisation conditions offers direct insight into how the first generations of stars enriched and reionised the intergalactic medium \citep[e.g.][]{Stark2016, PerezMontero2021,Endsley2024,LLerena2025}.

However, the mechanisms that trigger and sustain the EELG starburst phase remain uncertain. It has long been debated whether the extreme star formation in low-mass systems originates from external interactions, such as tidal encounters or mergers, or from processes not linked to an ongoing interaction, such as the accretion of gas from the circumgalactic medium or self-regulated star formation cycles. Observationally, both scenarios find partial support. Several studies have shown that interactions can enhance star formation in low-mass galaxies \citep{Lambas2003, Ellison2008, Patton2013,Lassen2025}, yet many compact starbursts appear to evolve in isolation \citep{Noeske2001, Koulouridis2013}. For EELGs specifically, \citet{Amorin2015} found that about one-third of the $z$COSMOS \citep{Lilly2007} EELGs reside in pairs or groups, while more recent analyses of GPs and EELGs using the Multi Unit Spectroscopic Explorer \citep[MUSE;][]{Bacon2010} and JWST data indicate that many are isolated \citep{Laufman2022, Gupta2023}, as also found by \citet{Kimsey2024} in a follow-up of 18 GPs. More generally, the environment is known to affect the gas content, interstellar medium (ISM) properties, and the recent star formation history of galaxies \citep{Peng2010, Darvish2016, Sobral2022}, and this connection can vary considerably depending on both stellar mass and cosmic epoch \citep{Zhou2025, Toni2025}, justifying a statistical approach for extreme low-mass systems such as EELGs.

The arrival of the Javalambre Physics of the Accelerating Universe Astrophysical Survey (J-PAS; \citealt{Benitez2014,Bonoli_21}), with its contiguous narrow-band filter system providing low-resolution optical spectra and accurate photometric redshifts \citep{Hernan-Caballero_21} over a large area of the sky, enables a comprehensive statistical analysis of the environment of galaxies (see Sect. \ref{sec:2.1}). In particular, these capabilities are ideal for quantifying pair statistics and environmental dependencies for low-mass, strongly star-forming systems, such as EELGs. 

In this work, our main goal is to characterise the environment of a sample of 917 EELGs from the most recent J-PAS Internal Data Release (IDR), selected by \cite{AntonioG}. To this end, we use a probabilistic approach based on the Pair Probability Function \citep[PPF,][hereafter LS15]{LopezSanJuan2015}. By integrating the full redshift probability distributions and considering galaxy pairs within a given projected separation, we quantify the likelihood of physical associations between galaxies and evaluate whether the extreme star-forming phase in EELGs is associated with the presence of a detectable companion. This work presents the first statistical study of the environments of EELGs within a wide-field narrow-band survey, combining the depth and spatial coverage of J-PAS with a probabilistic treatment of photometric redshifts to derive robust pair fractions and environmental trends. 

The structure of this paper is as follows: Sect. \ref{sec:2} describes the J-PAS data and the selection of the EELGs and companion samples; Sect. \ref{sec:3} details the methodology and construction of the PPF; Sect. \ref{sec:4} presents the results on pair fractions and their dependence on the projected distance; Sect. \ref{sec:5} compares the physical properties of EELGs and their companions; and Sect. \ref{sec:6} summarises our conclusions. The cosmological parameters adopted are: $\Omega_\mathrm{M} = 0.3$, $\Omega_\mathrm{\Lambda} = 0.7$, and $H_\mathrm{0} = 70$ km s$^{-1}$ Mpc$^{-1}$. We assume a \cite{Chabrier2003} initial mass function (IMF) for the estimation of both star formation rate (SFR) and $M_{*}$, and we use the convention of positive EWs for emission lines.


\section{J-PAS and data selection} \label{sec:2}

\subsection{J-PAS survey}\label{sec:2.1}
\begin{figure*}
    \includegraphics[trim= 0cm 0cm 0cm 0cm,width=\textwidth]{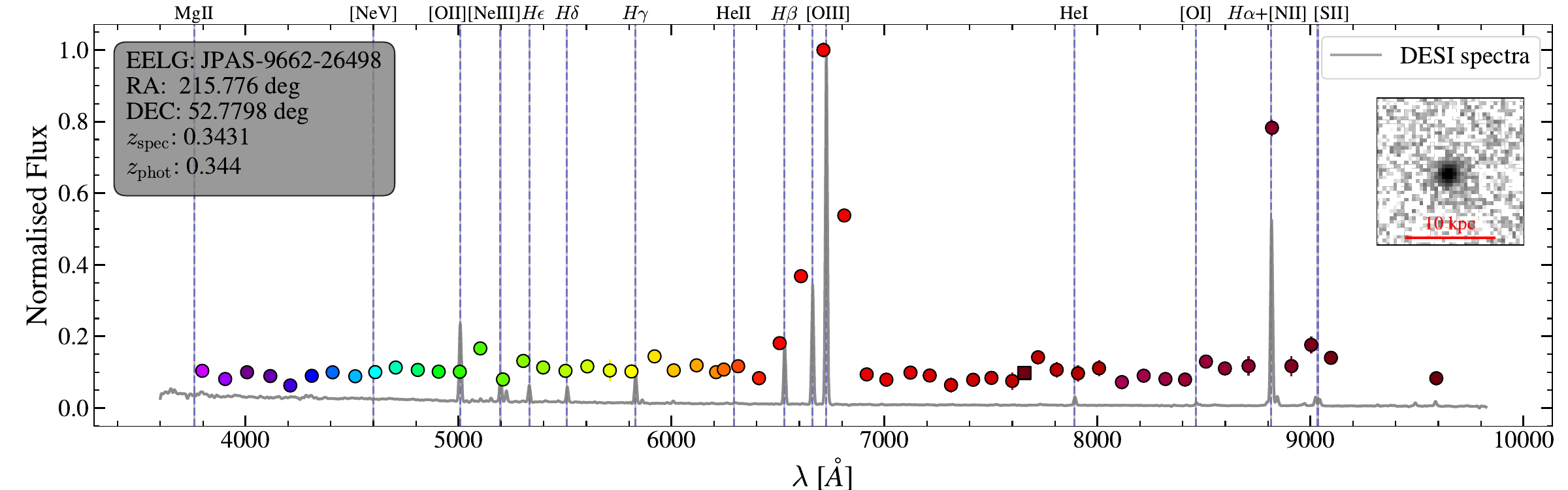}
    \caption{Example of the J-spectrum obtained from J-PAS for one of the EELGs (JPAS-9662-26498), at redshift $z\sim0.343$, studied in this work. The spectrum obtained for the same source in the DESI project, smoothed with a Gaussian profile for visualisation purposes, is shown in grey. Both spectra are normalised to the flux of the [\ion{O}{III}]$\,\lambda\,5007$ line. Strong emission lines are visible in both spectra. The dark red square at $\sim 7600\,\AA$ corresponds to the flux of the \textit{i}-SDSS band. The inset shows the J-PAS cut-out of the EELG.}
    \label{fig:J_spectra}
\end{figure*}

The J-PAS project\footnote{\url{https://www.j-pas.org/}} is a spectrophotometric survey aimed at scanning thousands of square degrees of the northern sky, using a unique filter system. It consists of 54 narrow-band filters that cover the entire optical range of the electromagnetic spectrum, from 3780\,\AA\ to 9100\,\AA. These narrow-band filters are spaced by about 100\,\AA, and have a full width at half maximum (FWHM) of $\sim$145\,\AA. In addition, there are also two mid-band filters, one in the near-UV regime and one in the near-IR range, and a broadband filter ($i$-SDSS; see figure 2 of \citealt{Bonoli_21} for more details). As a result, a low-resolution ($R \sim 60$) spectrum (hereafter J-spectrum) is obtained for each observed object. Figure \ref{fig:J_spectra} presents an example of the J-spectrum of one of the EELGs studied. For comparison, the spectrum of the same object obtained from DESI DR1 \citep{DESI2025} is shown, both normalised to the flux of the $[\ion{O}{III}]\,\lambda5007 $ line. The strong H$\alpha$, H$\beta$, and [\ion{O}{III}] emission lines are visible in both spectra. 

The J-PAS survey is designed to achieve a 5$\sigma$ limiting magnitude in the AB system of $m_{\mathrm{AB}}\sim 23.5$ in the broadband detection image and $m_{\mathrm{AB}} \sim 22.5$ for the case of narrowband detection, for point-like sources measured within a fixed aperture \citep{Benitez2014}. Observations are carried out at the Observatorio Astrofísico de Javalambre\footnote{\url{https://oajweb.cefca.es/}} using the 2.55\,m telescope (T250; \citealt{Cenarro2018}), equipped with the Javalambre Panoramic Camera (JPCam; \citealt{Taylor2014}), a wide-field 14-CCD mosaic camera, installed at the Cassegrain focus. JPCam covers approximately 5\,deg$^2$ with a pixel scale of 0.2267\,arcsec/pixel. This configuration is specifically designed for spectrophotometric surveys, allowing us to reach a photometric redshift precision of $\sigma_z/(1+z) \approx 0.003$ \citep{Hernan-Caballero_21,Hernan2023,Hernan2024}.

\subsection{Data selection}\label{sec:data_selec}
This study uses data from the J-PAS Internal Data Release IDR-202406 (hereafter IDR). The IDR includes data for the first $\sim$30\,deg$^2$ observed with the complete set of J-PAS filters. These data include positions, fluxes (J-spectra), and magnitudes derived with \texttt{SExtractor} \citep{Bertin1996}, and photometric redshift information, obtained through a modified version of \texttt{LePhare} \citep{Arnouts2011}, for approximately 850k sources up to $i_{\rm SDSS}<22.5 \,\mathrm{mag}$. In addition to this region, the IDR also includes observations over a much larger footprint of $\sim130\,\mathrm{deg}^2$, where sources have measurements in at least 20 J-PAS bands (and $i_{\rm SDSS}<22.5\,\mathrm{mag}$), amounting to roughly $3.5$ million objects. 

The IDR contains the probability distribution function (PDF) of the redshift as part of the photometric redshift information. This probability distribution is the posterior provided by \texttt{LePhare} and facilitates the determination of the probability that a given galaxy, $i$, lies within a specified range of redshifts ($z_1 < z < z_2$). This probability may be expressed as 
\begin{equation}
    P_i(z_1,z_2) = \int_{z_1}^{z_2} \mathrm{PDF}_i(z)\,dz \label{eq:PDFz}.
\end{equation}
The use of the complete photometric redshift PDFs allows the redshift uncertainties to be treated statistically, in particular through the overlap of the PDFs of two distinct galaxies. This process enables the calculation of the probability of two galaxies being at the same redshift, thus identifying potential companions. This is discussed in depth in Sect. \ref{sec:3}.

\subsubsection{The EELGs sample}
\begin{figure*}
    \includegraphics[trim= 0cm 0cm 0cm 0cm,width=\textwidth]{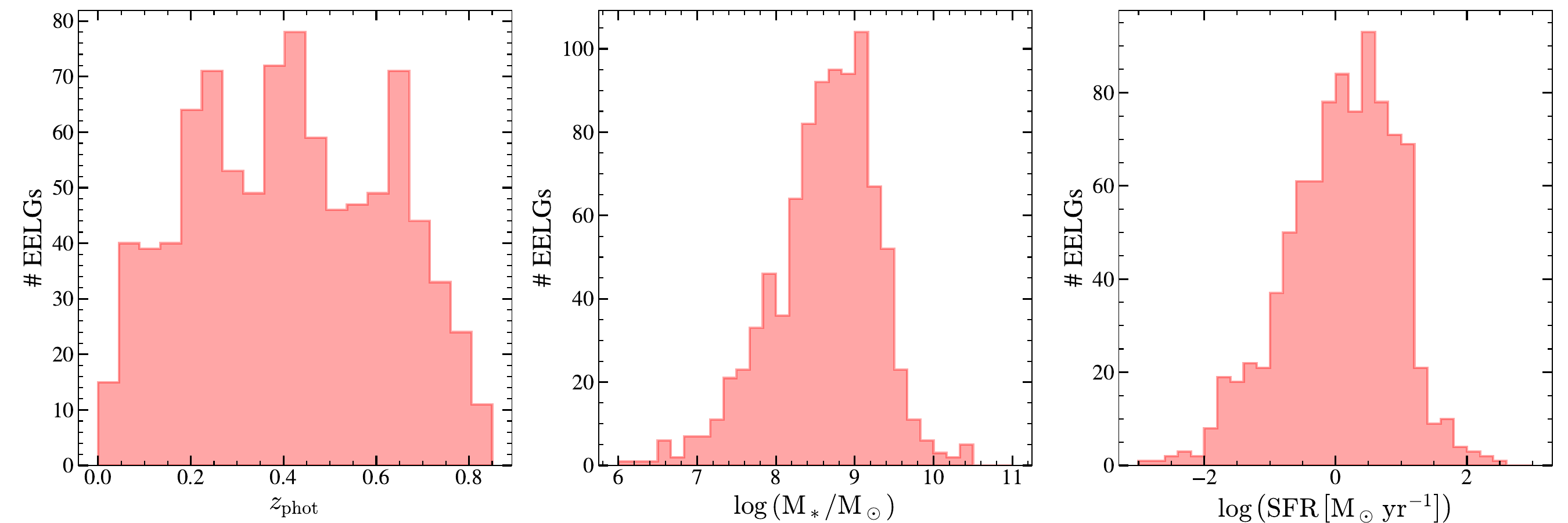}
    \caption{Distributions obtained for the photometric redshift (left), stellar mass (centre), and SFR (right) of the J-PAS EELGs.}
    \label{fig:EELGs}
\end{figure*}

The present study focuses on the analysis of EELGs detected in the J-PAS IDR catalogue. The selection was carried out in \cite{AntonioG}. We refer the reader to that paper for a detailed explanation. In summary, the sample was constructed following a two-step approach. First, EELGs were identified using a photometric contrast method applied to J-PAS narrow-band data, designed to detect strong emission lines with high EW ($>300\,\mathrm{\AA}$), following the methodology of \cite{Iglesias2022}. Subsequently, machine-learning techniques were used to remove contaminants and refine the classification, resulting in a final sample of 917 EELGs up to $z\sim0.9$ in approximately $30\,\mathrm{deg}^2$, with a purity of 95\% and an estimated AGN contamination of 6\%. These estimates are based on a subset of 79 sources with available spectroscopic counterparts in DESI DR1 and may therefore be subject to statistical uncertainties.

The photometric redshifts of the EELGs (see the left panel in Fig. \ref{fig:EELGs}) were also derived in \cite{AntonioG} by fitting the spectral energy distributions (SEDs) using a dedicated set of EELG templates, specifically designed to account for their strong emission lines and extreme spectral features. The resulting redshifts show excellent precision compared to the available spectroscopic redshifts (see figure 4 of \citealt{AntonioG}), with $\sigma_{\rm NMAD}\sim0.0015$, corresponding to a typical velocity uncertainty of $\sim450\,\mathrm{km\,s^{-1}}$. This level of precision is significantly better than the typical J-PAS photometric redshift accuracy ($\sim0.003$).

The physical properties of these sources were further characterised by fitting the SED with the Code Investigating GALaxy Emission software \citep[\texttt{CIGALE};][]{Boquien2019}, which models stellar populations, nebular emission, and dust attenuation using a Bayesian approach. The SED fittings used all the available photometric information of the EELGs in addition to the J-PAS fluxes, covering the electromagnetic spectrum from the FUV to the FIR range (see table C.1 of \citealt{AntonioG} for the \texttt{CIGALE} parameter settings). Most sources have derived stellar masses ranging from $10^7$ to $10^{10}\,\mathrm{M}_\odot$, with a median $\log(\mathrm{M}_*/\mathrm{M}_\odot) \sim8.68$. The SED fitting also provided SFR and accurate estimates of the ionising photon production efficiency, which is comparable to that expected at $z > 6$ \citep{AntonioG}. In Fig. \ref{fig:EELGs}, the distributions of the stellar mass and SFR derived for the selected EELGs are shown in the central and right panels, respectively. For comparison with their companions presented in Sect.~\ref{sec:5}, the EELGs were re-fitted with the same configuration as the companions (Sect.~\ref{sec:comp_sel}).

\subsubsection{Selection of companion candidates} \label{sec:comp_sel}
Following the creation of a catalogue of EELGs, the next step involves exploring the potential existence of companion galaxies in areas near the EELGs. To do this, an initial search is performed in the full J-PAS IDR catalogue ($\sim$130\,deg$^2$). To ensure the completeness of the sample of potential companion galaxies, the search in the IDR catalogue is only restricted by coordinates. The goal is to identify galaxies within a predetermined angular separation from each EELG. 
The preliminary search uses a maximum separation of 5\,arcmin, considered adequate to identify all potential companion galaxies associated with the EELGs, including those at low redshifts ($z<0.1$), which require larger angular search radii. 
In Sect. \ref{sec:3}, this initial coordinate-based selection is refined through the application of photometric-redshift and probability-based criteria, and additional quality filters, allowing us to distinguish high probability companions from random projected neighbours.

To study the galaxy pairs in detail, we fitted the SEDs of galaxies that are candidates to be companions of the EELGs (see Sect. \ref{sec:3}). To do this, we used \texttt{CIGALE} following the parameter settings of \cite{AntonioG}. However, these settings were slightly modified to also account for early-type galaxies and the AGN contribution. These settings are listed in Table \ref{tab:CIGALE}.
To ensure a homogeneous comparison between both members of each pair, the EELGs with companions were re-fitted with exactly the same configuration and photometric information as their companions. The stellar masses, SFRs, and metallicity estimates obtained are consistent with those of \cite{AntonioG}, whereas E(B--V) is systematically higher, reflecting the different attenuation prescriptions adopted in both works.


\section{Companion galaxy selection} \label{sec:3}
This section describes the analysis performed to identify the EELGs that have companion galaxies. We employ the approach defined by \citetalias{LopezSanJuan2015}, which uses the complete photometric redshift probability distribution instead of only the peak of the distribution, which is used to assign the nominal value of the photometric redshift of that source. This method allows for a more robust treatment of the associated uncertainties. Although the selection of potential companions follows that of \citetalias{LopezSanJuan2015}, it includes some specific modifications, so a comprehensive description is provided below.

\subsection{Selection criteria for companion candidates}\label{sec:companions}
When studying whether two galaxies can be considered companions, it is essential that they fulfil a particular separation criterion in order to qualify as a potential companion pair. Specifically, the projected separation on the plane of the sky ($r_p$) between them should fall within a specified interval. This interval is defined as $r_\mathrm{p}^{\rm min} < r_p < r_\mathrm{p}^{\rm max}$, where $r_\mathrm{p}^{\rm min}$ and $r_\mathrm{p}^{\rm max}$ represent the lower and upper bounds, respectively. To avoid selecting outer regions of the 
reference EELG
as possible companions, a minimum separation distance has to be imposed. We used a distance of 5$\,$kpc as the lower bound\footnote{The reference galaxy is always an EELG, with a typical size of up to $1-1.5\,\mathrm{kpc}$ \citep{Cardamone2009}. This ensures that we capture the entire reference EELG while avoiding the detection of outer parts as possible companions.}. For the upper bound, we adopted the value of 200$\,$kpc, identified by \cite{Cerdosino2024} as the optimal limit from the analysis of the completeness and contamination for different values of $r^{\rm max}_p$ in \textit{The Next Generation} Illustris \citep{Nelson2019} magnetohydrodynamical cosmological simulations.

On the other hand, the companion galaxies also have to be close along the line of sight. This criterion is usually expressed in terms of the line-of-sight velocity difference, setting a maximum value ($\Delta v_\mathrm{max}$). In the present study, we used $\Delta v_\mathrm{max} = 500\,\mathrm{km\,s^{-1}}$, which is close to the velocity resolution of J-PAS ($\Delta v=500\,\mathrm{km\,s^{-1}}$ at $z=0.2$). Furthermore, this value is frequently adopted in the field for selecting galaxy pairs (\citealt{Ellison2008}; \citetalias{LopezSanJuan2015}) because it offers a good balance between completeness and sample contamination \citep{Patton1997,Cerdosino2024}. The robustness of this choice is also tested in Sect.~\ref{sec:4.1} by recomputing the pair fraction using $\Delta v_{\rm max}=700$ and $900\,\mathrm{km\,s^{-1}}$.

Finally, when selecting possible companion galaxies for this study, we require that one member of each pair be an EELG. In our analysis, the EELG
is taken as the reference galaxy of the pair.
With this, we can obtain the number of companions of EELGs which, compared to the total number of EELGs, would give us the fraction of EELGs with companion galaxies ($f_\mathrm{EELGs\text{-}pair}$), that is,
\begin{equation} \label{eq:frac_pair}
    f_\mathrm{EELGs-pair} = \frac{N_\mathrm{pair}}{N_\mathrm{EELGs}},
\end{equation}
where $N_\mathrm{pair}$ is the number of galaxy pairs found and $N_\mathrm{EELGs}$ is the total number of EELGs in our initial sample.

\subsection{Pair Probability Function}
In this section, the methodology used to apply the conditions imposed for the selection of pairs of companion galaxies is outlined. 

\subsubsection{Line of sight selection}\label{sec:redshift}
\begin{figure*}
    \includegraphics[trim= 0cm 0cm 0cm 0cm,width=\textwidth]{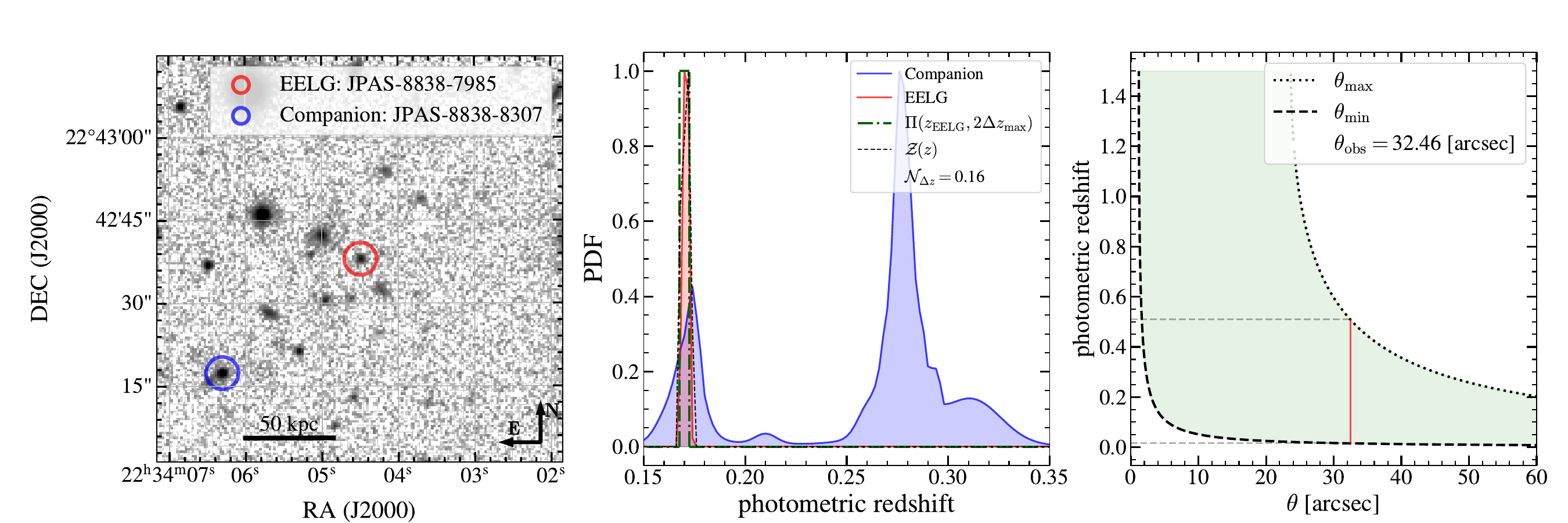}
    \caption{\textit{Left panel:} an example of a cut-out of the J-PAS field image in the $i$-SDSS band. The EELG JPAS-8838-7985 (red circle) is at the centre, and a possible companion, JPAS-8838-8307 (blue circle), lies to the south-east. The observed angular separation of these two galaxies is 32.46$\,$arcsec ($\sim 93\,\mathrm{kpc}$). \textit{Central panel:} PDFs of the photometric redshifts for the EELG, in red, and the potential companion galaxy, in blue, are shown. The black dashed line represents the redshift probability function, and the green dash-dotted line represents the top-hat function. The value of the integral of their product is shown in the legend (see text for details). \textit{Right panel:} the green shaded area in the diagram delineates the region in which the angular separation criteria imposed to select a galaxy as a possible companion are fulfilled, going from $\theta_\mathrm{min}$ to $\theta_\mathrm{max}$. This region marks the redshift range that is masked. For this galaxy pair, a separation of 32.46 arcsec (red line) is observed, so the galaxies with redshifts ranging from 0.02 to 0.51 (horizontal grey shaded lines) fulfil the minimum and maximum separation criteria of 5$\,$ and 200$\,$kpc, respectively.}
    \label{fig:pdf_sample}
\end{figure*}

This study begins with an analysis of the EELGs catalogue along with the associated catalogue of potential companion galaxies. The proximity of these galaxies along the line of sight is subsequently evaluated using photometric redshift data. To achieve this goal, the comprehensive redshift PDF, as described in Eq. \ref{eq:PDFz}, is used for both the EELG and its potential companion galaxy.

An EELG may have several potential companions. In these situations, where more than one potential companion is present, the analysis is performed per pair. Consequently, for every individual galaxy pair, the redshift probability function, denoted as $\mathcal{Z}(z)$, is constructed, and it is specified as follows:
\begin{equation}
    \mathcal{Z}(z) = \frac{2 \times \mathrm{PDF}_1(z) \times \mathrm{PDF}_2(z)}{\mathrm{PDF}_1(z)+\mathrm{PDF}_2(z)} = \frac{\mathrm{PDF}_1(z) \times \mathrm{PDF}_2(z)}{N(z)}
\end{equation}
where $N(z)$ serves as a normalisation factor, referring to the estimated potential number of galaxy pairs at a specific redshift $z$, and this factor is determined by
\begin{equation}
    N(z) = \frac{\mathrm{PDF}_1(z)+\mathrm{PDF}_2(z)}{2} 
\end{equation}
and, by construction, it satisfies $\int N(z)\,dz = 1$. Thus, $\mathcal{Z}(z)$ represents the fractional number of pairs at redshift $z$. Consequently, the integral of $\mathcal{Z}(z)$ over the entire redshift range is
\begin{equation}
    \mathcal{N}_z = \int \mathcal{Z}(z)\,dz,
\end{equation}
which returns the fraction of pairs for the two-galaxy system. This value ranges from 0 to 1. The quantity $\mathcal{N}_z$ represents the probability that the system of two galaxies is at the same redshift, based on the probabilistic information of the photometric redshifts, which in turn takes into account the photometric redshift error associated with each galaxy.

In order to consider the imposed criterion for pair selection, it is necessary to impose a limit on the velocity difference between the two galaxies ($\Delta v_\mathrm{max}=500\,\mathrm{km\,s^{-1}}$). This difference in velocity can be translated directly into the maximum difference in the redshift space ($\Delta z_\mathrm{max}$) using the following expression
\begin{equation}
    \Delta z_\mathrm{max} = \frac{\Delta v_\mathrm{max}\times (1+z_\mathrm{EELG})}{c},
\end{equation}
where $z_\mathrm{EELG}$ is the redshift of the reference EELG and $c$ is the speed of light. Consequently, to appropriately incorporate this factor into the analysis, the redshift probability function is multiplied by a top-hat function. This is mathematically represented by 
\begin{equation}
    \mathcal{N}_{\Delta z} = \int \Pi(z_\mathrm{EELG},2\Delta z_\mathrm{max})\times\mathcal{Z}(z)\,dz ,
\end{equation}
where $\Pi$ is a top-hat function of width $2\Delta z_\mathrm{max}$ centred at the redshift of the EELG ($z_\mathrm{EELG}$). Thus, $\mathcal{N}_{\Delta z}$ represents the fractional number of pairs that meet the criterion of $\Delta v_\mathrm{max}=500\,\mathrm{km\,s^{-1}}$.
    
The left panel of Fig. \ref{fig:pdf_sample} shows a real example of an EELG and a companion candidate, and in the central panel of Fig. \ref{fig:pdf_sample}, the PDFs of the photometric redshifts for the pair are presented. Moreover, the redshift probability function, the top-hat function, and the integral of their product, representing the total fraction of pairs, are shown. The method is most effective when dealing with complex shapes of the PDFs, such as that of the companion galaxy, because it uses all the photometric redshift information, as opposed to merely the peak of the distribution.

\subsubsection{Projected separation on the sky}\label{sec:theta}
The next step in the search for pairs of companion galaxies is to analyse the angular separation between the objects. To this end, we searched for galaxies that satisfy the angular separation criterion in the J-PAS database for each EELG.

The minimum and maximum projected separations are converted into angular separations  using cosmological principles. That is, for the values $r_\mathrm{p}^\mathrm{min}$ and $r_{\rm p}^\mathrm{max}$, this transformation process is
\begin{equation}
    \theta_\mathrm{min}(z) = \frac{r_\mathrm{p}^\mathrm{min}}{d_\mathrm{A}(z)}
\end{equation}
and
\begin{equation}
    \theta_\mathrm{max}(z) = \frac{r_{\rm p}^\mathrm{max}}{d_\mathrm{A}(z)},
\end{equation}
where $d_\mathrm{A}(z)$ denotes the angular diameter distance corresponding to a given redshift $z$. Formally, this concept is expressed as a selection mask, which is assigned a value of 1 if the specified criterion is satisfied at a given redshift and a value of 0 otherwise. Therefore, the selection mask is given by
\begin{equation}
    \mathcal{M}^\theta(z) = 
    \begin{cases}
      1, & \text{if}\ \; \theta_\mathrm{min}(z) \leq \theta \leq \theta_\mathrm{max}(z)\\
      0, & \text{otherwise.}
    \end{cases}
\end{equation}

The green shaded area in the right panel of Fig. \ref{fig:pdf_sample} marks the region where the angular separation criterion is fulfilled for the redshift range studied in this work. For the galaxy pair shown in the left panel of Fig. \ref{fig:pdf_sample}, where the observed angular separation is $\theta\sim32\,\mathrm{arcsec}$ ($\sim 93\,\mathrm{kpc}$), the redshift range that satisfies the minimum and maximum separations is $0.02\lesssim z\lesssim0.51$.

\subsubsection{Mass completeness}\label{sec:mass_comp}
\begin{figure}
    \includegraphics[trim= 0cm 0cm 0cm 0cm,width=\columnwidth]{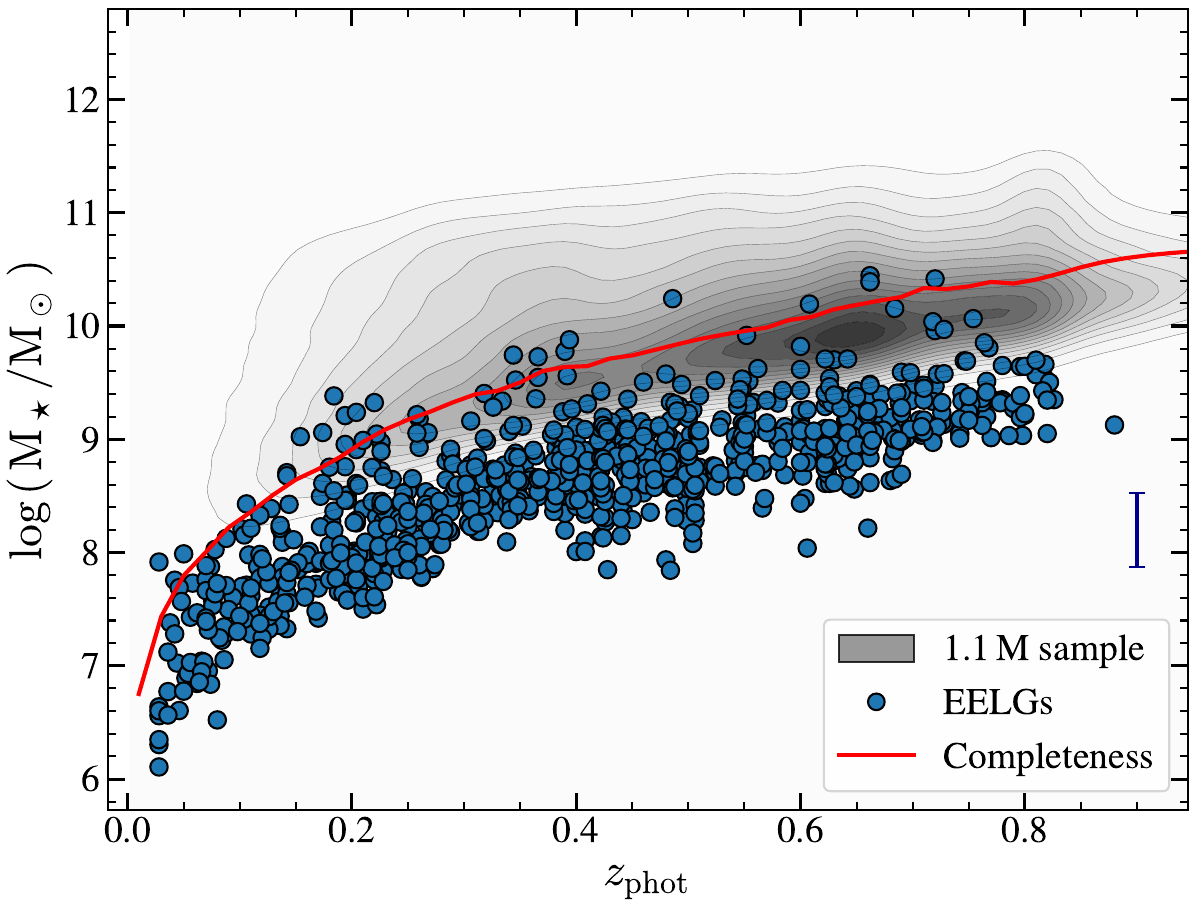}
    \caption{Stellar mass as a function of photometric redshift. The 90\% mass completeness for the J-PAS survey is illustrated by the red line, derived using a $1.1\,$M sample subset (grey density contours) randomly chosen from the catalogue. The blue dots denote the EELGs analysed in this study. Additionally, the blue error bar shows the median uncertainty associated with the stellar mass of these EELGs.}
    \label{fig:mass_completeness}
\end{figure}

The statistical characterisation of galaxy environments requires a well-defined selection function to avoid systematic biases. In the current J-PAS IDR, photometric redshifts and their associated PDFs are computed only for sources with $i_{\mathrm{AB}} < 22.5$. Since our environmental analysis relies on the PPF, which requires well-sampled $\mathrm{PDFs}$ for both members of a potential pair, the effective depth of the companion search is therefore limited by the magnitude range in which photometric redshifts are available for the entire galaxy population.

This magnitude threshold does not correspond to the photometric completeness of J-PAS, which varies from filter to filter in the range $m_{\mathrm{AB}} \sim 22 - 23\,\mathrm{mag}$ (Díaz-García et al., in prep.), but rather to the operational limit adopted for the computation of photometric redshifts. Therefore, we adopt a representative value of $i_{\mathrm{lim}} = 22.5$ to characterise the effective selection limit of our sample. Restricting the environmental analysis to galaxies brighter than this threshold ensures that all potential neighbours can be assigned a reliable redshift probability distribution, avoiding magnitude-dependent biases in the derivation of pair fractions.

To estimate the stellar mass completeness of the J-PAS survey, we used a sample of 1.1$\,$M galaxies with photometric redshifts (Vázquez-Ramió et al., in prep.) and stellar masses (González-Delgado et al., in prep.) from the J-PAS IDR catalogue. The J-spectra were fitted using the Bayesian SED-fitting code \texttt{BaySeAGal} \citep{Rosa2021}, assuming a delayed-$\tau$ star formation history (SFH). Stellar masses are then derived from the total mass converted into stars according to the inferred SFH and the observed galaxy luminosity, accounting for mass loss due to stellar evolution, as predicted by single stellar population (SSP) models.

We adopt the formalism from \citet{Pozzetti2010} to translate the observational limit into a stellar mass completeness threshold of 90\%. Although this formalism is strictly valid for flux-limited samples that are complete in detection, it has been shown to provide a reasonable approximation to the stellar mass completeness in J-PAS-like datasets (e.g. miniJPAS; \citealt{DiazGarcia2024}). In this work, the method is applied to characterise the effective mass completeness of the galaxy sample with reliable photometric redshifts. In order to carry out this process, galaxies were divided into redshift bins $\Delta z=0.02$, and the faintest 20\% of the galaxies were selected. For each of these galaxies, we scaled their mass as if they had an apparent magnitude equal to the survey limit. Thus, for each galaxy at a given redshift, the limiting stellar mass is defined as follows:
\begin{equation}
\log \mathrm{M}_{*}^{\mathrm{lim}}(z) = \log \mathrm{M}_{*} + 0.4 \times\left[\, i_\mathrm{AB}(z) - i_{\mathrm{lim}} \,\right],
\end{equation}
where $\mathrm{M}_{*}$ is the stellar mass of the galaxy, $i_\mathrm{AB}(z)$ is the observed $i$-band magnitude, and $i_{\mathrm{lim}}=22.5$ is the completeness limit of J-PAS. Finally, the completeness of the mass at 90\% $(\,\mathrm{M}^\mathrm{comp}_*\,)$ is selected as the 90th percentile of the scaled mass distribution for each redshift bin.
In Fig. \ref{fig:mass_completeness}, where the stellar mass is shown against the photometric redshift, we can see the 90\% mass completeness obtained for the J-PAS survey for each redshift.

In practice, we apply a completeness mask $\mathcal{M}^\mathrm{mass}(z)$ such that:
\begin{equation}
    \mathcal{M}^\mathrm{mass}(z) = 
    \begin{cases} 
        1 & \text{if } \,\mathrm{M_*} \geq \mathrm{M}^\mathrm{comp}_*(z), \\
        0 & \text{if } \, \mathrm{M_*} < \mathrm{M}^\mathrm{comp}_*(z)
    \end{cases}
\end{equation}
where $\mathrm{M_*}$ is the stellar mass of the galaxy and $\mathrm{M}^\mathrm{comp}_*(z)$ is the 90\% completeness at the redshift of the galaxy.

\subsubsection{Pair Probability Function construction}\label{sec:PPF}
We are now able to construct a function that enables us to identify galaxy pairs probabilistically, using the PDF of the photometric redshift. The PPF is constructed as the product of all the factors described above:
\begin{equation}\label{eq:PPF}
    \mathrm{PPF}(z) = \mathcal{Z}(z) \times \Pi (z_\mathrm{EELG},\,2\Delta z_\mathrm{max}) \times  \mathcal{M^\theta}(z) \times \mathcal{M^\mathrm{mass}}(z),
\end{equation}
where $\mathcal{Z}$ is the redshift probability function (Sect. \ref{sec:redshift}), $\Pi$ is the top-hat function centred at the redshift of the EELG in the pair and with a width of twice the maximum redshift difference (Sect. \ref{sec:redshift}), $\mathcal{M^\theta}$ is the angular mask used to apply the minimum and maximum separation criteria in the plane of the sky (Sect. \ref{sec:theta}), and $\mathcal{M^\mathrm{mass}}$ is the mask to limit the companion sample to the stellar mass regime where J-PAS is complete at the 90\% level (Sect. \ref{sec:mass_comp}).
Thus, the integral of the PPF in the redshift range for a galaxy pair $i$,
\begin{equation}\label{eq:PPF_int}
    \int_{z_\mathrm{min}}^{z_\mathrm{max}} \mathrm{PPF_i}(z)\,dz
\end{equation}
gives the fractional number of pairs associated with the pair $i$ in that redshift range.

\subsection{Corrections for selection effects}\label{sec:error}
The identification of galaxy companions and the derivation of pair statistics are affected by several observational and selection effects. In this section, we describe the corrections applied to account for photometric redshift quality and survey observational effects.
\subsubsection{Photometric redshift \texttt{odds} analysis}\label{sec:error_z}
One factor that must be considered in the selection process is the quality of the photometric redshifts. \citetalias{LopezSanJuan2015} introduced the \texttt{odds} sampling rate (OSR) to address this issue, which has been widely adopted \citep[among others]{Mundy2017,Conselice2022,Puskas2025}.

The foundation of this concept is the \texttt{odds} quality parameter, $\mathcal{O}$, which evaluates the reliability of photometric redshifts \citep{Benitez2000, Molino2014}. For a galaxy $j$, the \texttt{odds} parameter is defined as the integral of the photometric redshift PDF over a narrow region around the peak, $z_\mathrm{b}$, given by
\begin{equation}
    \mathcal{O}_j = \int^{+\mathcal{K}(1+z_\mathrm{b})}_{-\mathcal{K}(1+z_\mathrm{b})} \mathrm{PDF_j}(z)\,dz
\end{equation}
where $\mathcal{K}$ represents the width of the interval around the peak and varies with each survey. Specifically, for the J-PAS survey, it was established $\mathcal{K} = 0.03$ \citep{Hernan-Caballero_21}.

Selecting objects based on their $\mathcal{O}$ value ensures high-quality photometric redshifts and reduces the incidence of outliers \citep{Molino2014}. We adopted $\mathcal{O}\geq0.3$ as the selection threshold for both EELGs and companion galaxy candidates in this study, a value commonly used in the literature (\citetalias{LopezSanJuan2015}; \citealt{Mundy2017,Conselice2022,Puskas2025}). Note that all EELGs meet this criterion, underscoring the accuracy of the photometric redshift estimates.

As defined by \citetalias{LopezSanJuan2015}, the OSR represents the ratio of galaxies that meet the criterion to the total galaxy count in the sample:
\begin{equation}
    \mathrm{OSR}(m) = \frac{\sum N(\mathcal{O}\geq0.3)}{\sum N(\mathcal{O}\geq0)}
\end{equation}
The OSR depends only on the magnitude $m$ of each object. Consequently, the OSR function for the J-PAS survey was formulated by using all catalogued objects with photometric redshift. We used the magnitude in the $i$-SDSS band with binning $\Delta m=0.2$ and used linear interpolation to determine the OSR at any magnitude (Fig. \ref{fig:OSR}). The correction factor for a galaxy $j$ is thus the inverse of its OSR:
\begin{equation}\label{eq:OSR}
    w^{\mathrm{OSR}}_j = \frac{1}{\mathrm{OSR}(m_j)}
\end{equation}
where $m_j$ is its magnitude in the $i$-SDSS photometric band.

\subsubsection{Survey edges and artefacts}\label{sec:area}
\begin{figure}
    \includegraphics[trim= 1cm 0.4cm 1cm 0cm,width=\columnwidth]{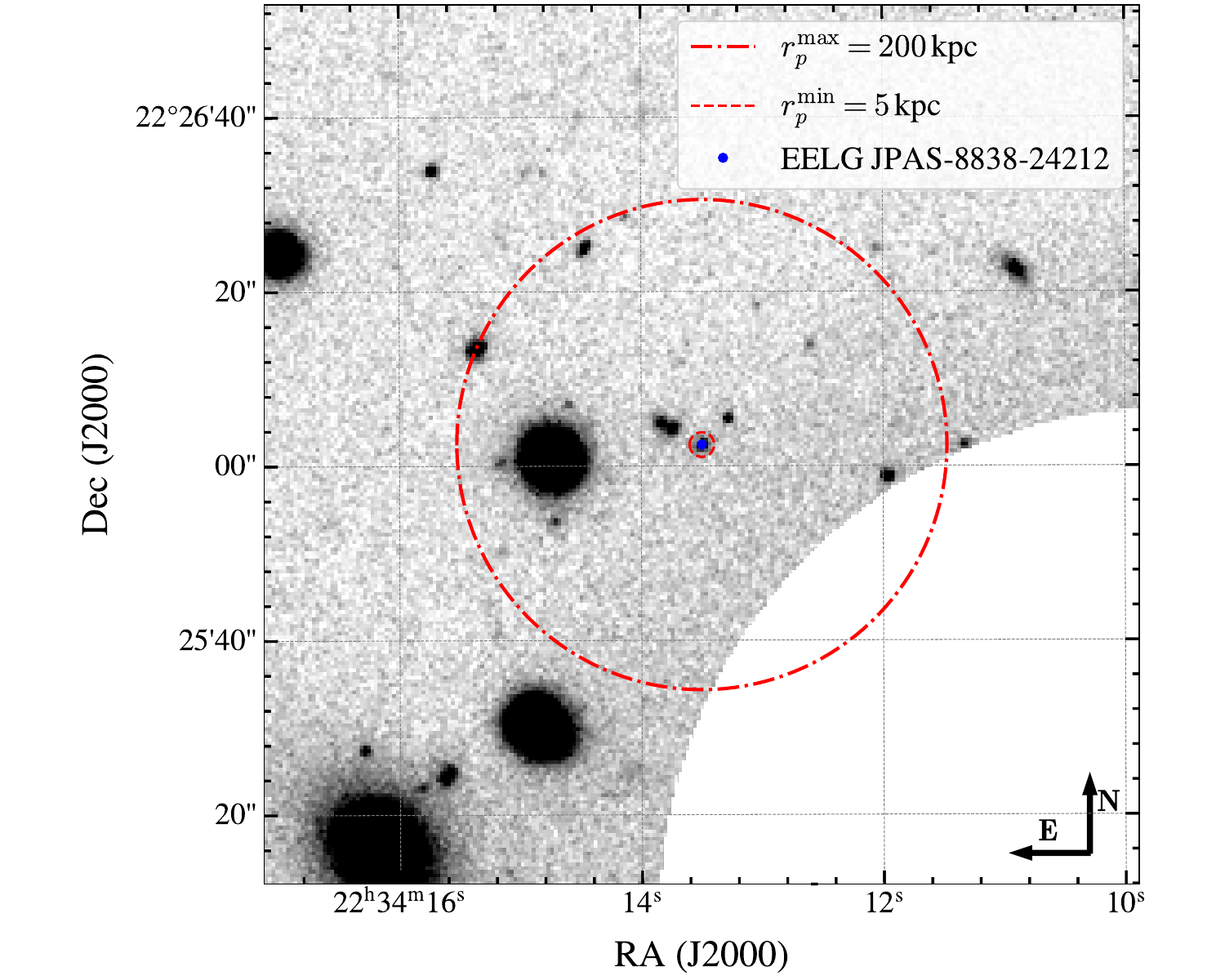}
    \caption{Example of the loss of the search area for companion galaxies around the reference EELG (blue dot) at $z=0.646$. In this example, the missing area is due to the presence of a bright star to the south-west of the EELG. The red dashed circle marks the minimum search value for the projected distance in the plane of the sky, $r_p^{\rm min}=5\,\mathrm{kpc}$, and the red dash-dotted circle marks the maximum search value, $r_p^{\rm max}=200\,\mathrm{kpc}$. In this case, the correction factor for the missing area is $w^\mathrm{mask}(z=0.646)=1.12$, that is a missing area of 11\%.}
    \label{fig:lost_area}
\end{figure}
Another source of error in the search for companion galaxies is the effect of the edges of the observed fields. For example, if an EELG is located at the edge of a field, the area available for companion searches is smaller than for EELGs located in the centre of a field. Bright stars close to the EELGs have a similar effect, as saturated regions limit the search area.

To correct for this effect, the J-PAS catalogue contains masks of the images, which trace all such artefacts. Using these masks, we determine the percentage of the search area available for each EELG, that is, the fraction of the area between $r_p^{\rm min}$ and $r_p^{\rm max}$ in which we can search for companion galaxies ($f_\mathrm{mask}$). Therefore, the correction factor is the inverse of the fraction of the area available for searching:
\begin{equation}\label{eq:area}
    w^{\mathrm{mask}}(z) = \frac{1}{f_\mathrm{mask}(z)}.
\end{equation}
It should be noted that $f_\mathrm{mask}$ depends on the redshift, as $d_{\rm A}$ does. However, this correction factor depends only on the EELG, not on the candidates for companion galaxies. Figure \ref{fig:lost_area} shows an example of the masked area at the redshift of the EELG. In this example, a bright star is observed to the south-west of the EELG, producing the masked region and lost area for searching for galaxy companions.

\subsection{Pair fraction for EELGs in J-PAS}\label{sec:selection}
Considering all the factors mentioned above, we can derive a function to calculate the number of galaxy pairs. 
Taking each EELG as the reference galaxy,
the number of pairs for each EELG, indexed by $i$, is expressed as:
\begin{equation}\label{eq:npair}
        N_\mathrm{pair}^{i} = \sum_j \int_{z_{\rm min}}^{z_{\rm max}} w^{\rm OSR}_\mathrm{j} \times w^{\rm mask}_i(z) \times \mathrm{PPF}_{i,j}(z)\,dz,
\end{equation}
where $j$ runs over all candidate companion galaxies of the reference EELG, $w^{\rm OSR}_\mathrm{j}$ is the correction factor based on the $\mathcal{O}$ parameter and $i$-SDSS band magnitude (Sect. \ref{sec:error_z}), $w^{\rm mask}$ is the correction factor to account for the lost area in the search for galaxy companions (Sect. \ref{sec:area}), and ${\rm PPF}_{i,j}$ represents the PPF for the pair (Eq. \ref{eq:PPF}). 

In addition, it is essential to correct for the influence of each EELG on the overall sample distribution. This correction, within the redshift range studied, is determined as follows:
\begin{equation}
    N_\mathrm{EELG}^{i} = \int_{z_{\rm min}}^{z_{\rm max}} {\rm PDF}_i (z)\, dz,
\end{equation}
where $i$ designates each EELG and PDF$(z)$ represents the probability distribution function of the photometric redshift (Eq. \ref{eq:PDFz}).

It is now possible to determine the proportion of EELGs that exhibit a companion galaxy that meets the specified criteria. This is called the pair fraction of the sample. To do this, we use a modified version of Eq. \ref{eq:frac_pair}, adjusted for the probabilistic approach outlined above. The resulting expression is
\begin{equation}\label{eq:f_pair}
    f_{\rm EELGs-pair} = \frac{\sum_i N_{\rm pair}^{i}}{\sum_i N_{\rm EELG}^{i}}.
\end{equation}
We calculate the error in the pair fraction and the final value using the bootstrap technique \citep{Efron1979}. By performing  $N_\mathrm{boots}=10^4$ iterations, we established a distribution of the pair fraction, where the mode of the distribution is adopted as the final value, and the 68\% interval around it as its uncertainty.

\section{Analysis of the pair fraction in EELGs} \label{sec:4}
As previously discussed, this research investigates whether companion galaxies could account for the significant star formation bursts observed in EELGs. To explore this hypothesis, we have devised a statistical method (see Eq. \ref{eq:npair}) that allows us to evaluate the presence or absence of companion galaxies in relation to EELGs. In this section, we outline how the pair fraction is corrected for random projections and explain how this fraction evolves with the maximum distance projected onto the sky plane.

\subsection{Randomness contribution to the pair fraction}\label{sec:4.1}
\begin{figure}
    \includegraphics[trim= 0cm 0cm 0cm 0cm,width=\columnwidth]{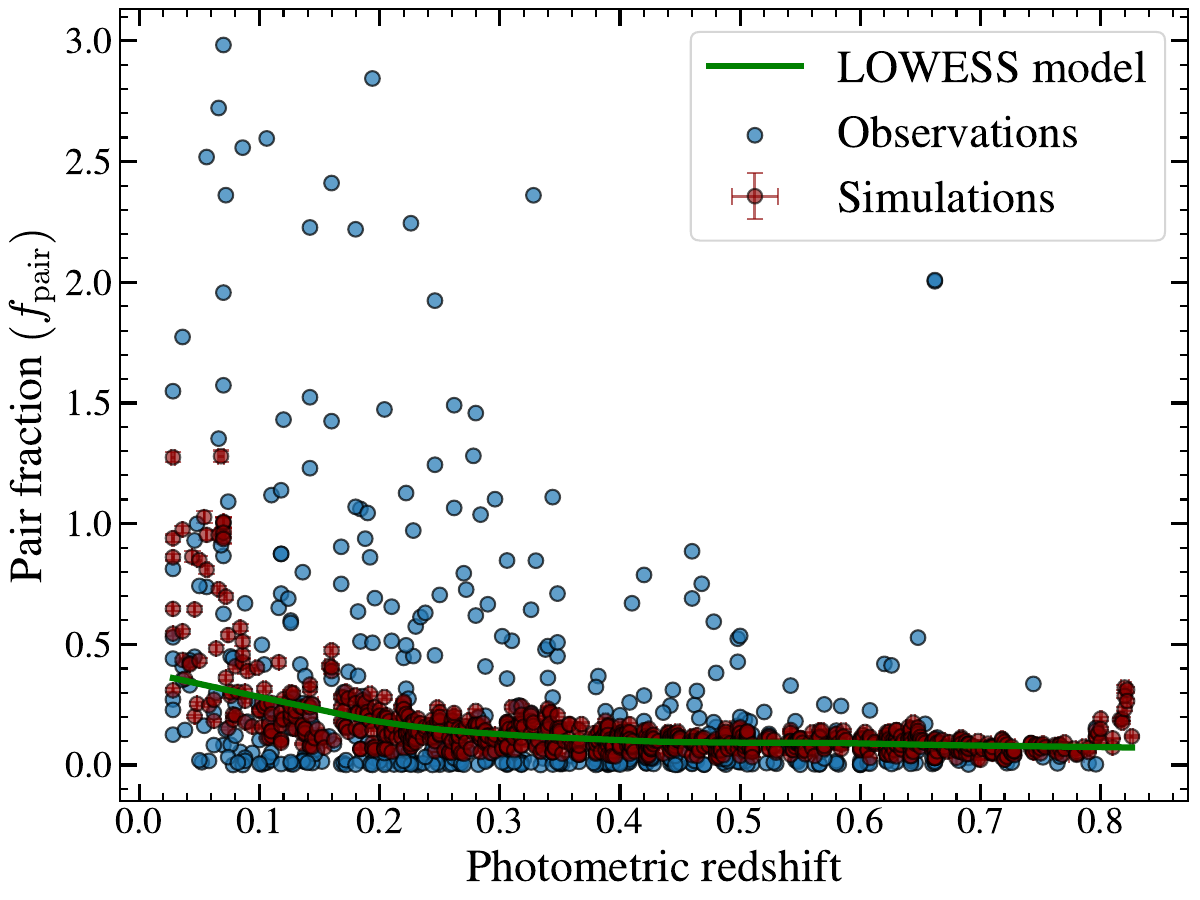}
    \caption{Pair fraction of each EELG against photometric redshift obtained for simulations at random positions in the J-PAS field (red dots) and obtained from J-PAS observations (blue dots). The empirical model of the simulation obtained using LOWESS method is represented by the green line.}
    \label{fig:lowees}
\end{figure}
In order to properly quantify the physical significance of the observed pair fraction for the EELG sample, it is crucial to estimate the fraction of projected associations expected by chance within the J-PAS survey. Even when restrictive criteria are imposed on projected distance and velocity separation, the width of the photometric redshift PDFs produces a non-negligible number of spurious alignments. This effect is particularly severe at low redshift, where the angular search area is larger, and may artificially enhance the measured pair fraction.

To assess this ``randomness contribution'', we performed a set of simulations in which each EELG was placed in $\sim$1300 random positions within 
the $30\,\mathrm{deg}^2$ area covered by the complete set of J-PAS filters, from which the EELG sample was selected, preserving the same survey geometry, masks, and completeness conditions. In each random realisation, the photometric redshift and PDF of the EELG were kept fixed, and only its sky position was changed. Potential companions around real and random positions are drawn from the same J-PAS IDR catalogue ($\sim130\,\mathrm{deg}^2$) and are subject to the same \texttt{odds} threshold ($\mathcal{O}\geq 0.3$) and OSR correction. 
For each random realisation, we recomputed the pair fraction following exactly the same procedure applied to the observed data. Finally, for each EELG we derived its random pair fraction ($f_\mathrm{pair}^\mathrm{rand}$) and its associated uncertainty using the same bootstrap technique as for the observed data, namely, we used  $N_\mathrm{boots}=10^4$ iterations to obtain a distribution for $f_\mathrm{pair}^\mathrm{rand}$ from which the value and error are assigned as the mode and 68\% interval around it, respectively.
The outcome of these simulations is a distribution of $f_{\mathrm{pair}}^{\mathrm{rand}}$ as a function of the redshift, which represents the contribution expected purely from chance projections. To correct for this effect, we modelled the $f_{\mathrm{pair}}^{\mathrm{rand}}$ as a continuous function of the redshift using a Locally Weighted Scatter-plot Smoothing (LOWESS) regression \citep{Cleveland79,Cleveland88}. LOWESS is a non-parametric, locally adaptive smoothing technique that fits low-order polynomials to subsets of the data, weighted by their proximity in the independent variable. This approach is particularly suitable in this context, as it does not require assuming an analytical form for the redshift dependence. Furthermore, Monte Carlo realisations were used in the LOWESS fit to account for the uncertainties associated with the simulations. Figure  \ref{fig:lowees} shows the pair fraction obtained from the simulations (red dots), along with the LOWESS model (green line). In addition, the values obtained from the observations for each EELG can also be seen (blue dots). The random contribution to the pair fraction closely mirrors the behaviour of the vast majority of objects. Only $\sim35\,\%$ of the observed pair fractions exceed the random contribution.

The resulting LOWESS curve provides a model for the random contribution to the pair fraction. Thus, the corrected observational pair fraction for an EELG, named $i$, is then obtained as:
\begin{equation}
f_{\mathrm{pair,\,}i}^{\mathrm{corr}}(z) = f_{\mathrm{pair,\,}i}^{\mathrm{obs}} - f_{\mathrm{pair,\,}i}^{\mathrm{rand}}(z),    
\end{equation}
where $f_{\mathrm{pair,\,}i}^{\mathrm{obs}}$ represents the pair fraction measured from the observations for an EELG $i$. This correction removes the expected contribution from random alignments while preserving the intrinsic redshift dependence of the physical pair fraction.
In cases where the corrected pair fraction was negative, the contribution to the pair fraction was set to zero. Furthermore, the uncertainty of the LOWESS model, estimated from Monte Carlo realisations, is below 0.005 ($\lesssim 1-2\,\%$) over the full redshift range, and its propagation into the final uncertainty of the pair fraction is negligible ($<1\,\%$).

After correcting for the random component, the pair fraction value obtained for the complete EELGs sample in J-PAS, following Eq. \ref{eq:f_pair}, is $f_\mathrm{EELGs\text{-}pair}=0.162^{+0.048}_{-0.033}$ for the predefined constraints of $\Delta v_{\rm max} = 500\,\mathrm{km\,s^{-1}}$ and $r_p^{\rm max} = 200\, \mathrm{kpc}$. This value is further supported by a consistency test performed using the available spectroscopic redshifts, which, despite the small number of objects, 
yields results consistent with our photometric redshift approach
(see Appendix \ref{sec:appendixC}).

Since the adopted velocity threshold may influence the estimated pair fraction when using photometric redshifts, we also explored the impact of varying the velocity window. We recomputed the pair fraction adopting thresholds of $\Delta v_\mathrm{max}=700$ and $900\,\mathrm{km\,s^{-1}}$, corresponding approximately to $\sim1.5$ and $\sim2$ times the characteristic photometric redshift uncertainty of the EELG sample. The resulting pair fractions as a function of projected separation are presented in Table~\ref{tab:pair_fraction_rp}. The pair fraction is observed to increase slightly when larger velocity windows are adopted; however, the change is modest and remains within the uncertainties for most separations. In particular, for $r_p^{\rm max}=200\,\mathrm{kpc}$ the pair fraction increases from $0.162^{+0.048}_{-0.033}$ for $\Delta v_\mathrm{max}=500\,\mathrm{km\,s^{-1}}$ to $0.195^{+0.042}_{-0.030}$ for $\Delta v_\mathrm{max}=900\,\mathrm{km\,s^{-1}}$, corresponding to an absolute increase of $\Delta f_\mathrm{pair}\sim0.033$ (a relative increase of $\sim20\,\%$). Most of this increase arises from low-redshift systems, where the adopted top-hat window may become narrower than the intrinsic width of the EELG photometric redshift PDFs, causing a small fraction of the probability distribution to fall outside the integration range. Nevertheless, as shown by the above comparison, the amount of probability lost in this way is limited and does not significantly affect the overall results.

The pair fraction value obtained in this study is lower than that reported by \cite{Amorin2015} for a sample of 165 EELGs in the zCOSMOS survey with redshifts ranging from 0.11 to 0.93. The authors found that $\sim30\,\%$ of the EELGs were in groups with one or more companion galaxies. In addition, this value is lower than that found in other studies for different types of galaxies. For example, \cite{Cerdosino2024} found a pair fraction value of $\sim 30\,\%$ for two galaxy samples from the S-PLUS survey \citep{Mendes2019}, within the redshift range $0<z<0.4$, $r$-band limiting magnitudes of 20.2 and 20.8 mag, and the criteria $\Delta v=1000\,\mathrm{km\,s^{-1}}$ and $r_p=150\,h^{-1}\,\mathrm{kpc}$. However, this discrepancy could be due to the different galaxy populations, as the authors studied more massive sources with a median stellar mass $\mathrm{M}_*\sim6\times10^{10}\mathrm{M}_\odot$.

\begin{table}
\setlength{\extrarowheight}{2pt}
\centering
\caption{Pair fraction as a function of projected separation $r_{\mathrm p}$ for different velocity thresholds.}
\label{tab:pair_fraction_rp}
\begin{tabular}{c c c c}
\hline\hline
$r_{\mathrm p}$  & $\Delta v_\mathrm{max} = 500$  & $\Delta v_\mathrm{max} = 700$  & $\Delta v_\mathrm{max} = 900$  \\
 (kpc)                &  (km\,s$^{-1}$) &  (km\,s$^{-1}$) &  (km\,s$^{-1}$) \\
\hline
30  & $0.015^{+0.004}_{-0.003}$ & $0.018^{+0.006}_{-0.003}$ & $0.018^{+0.006}_{-0.002}$ \\
50  & $0.033^{+0.008}_{-0.007}$ & $0.039^{+0.007}_{-0.008}$ & $0.039^{+0.008}_{-0.007}$ \\
75  & $0.046^{+0.007}_{-0.009}$ & $0.050^{+0.012}_{-0.006}$ & $0.054^{+0.009}_{-0.008}$ \\
100 & $0.063^{+0.013}_{-0.010}$ & $0.076^{+0.010}_{-0.014}$ & $0.072^{+0.015}_{-0.008}$ \\
125 & $0.090^{+0.025}_{-0.019}$ & $0.103^{+0.021}_{-0.018}$ & $0.107^{+0.024}_{-0.018}$ \\
150 & $0.108^{+0.044}_{-0.018}$ & $0.133^{+0.032}_{-0.027}$ & $0.136^{+0.037}_{-0.025}$ \\
175 & $0.136^{+0.034}_{-0.023}$ & $0.157^{+0.036}_{-0.026}$ & $0.166^{+0.038}_{-0.022}$ \\
200 & $0.162^{+0.048}_{-0.033}$ & $0.185^{+0.041}_{-0.029}$ & $0.195^{+0.042}_{-0.030}$ \\
\hline
\end{tabular}
\end{table}

\subsection{Evolution of the pair fraction with the projected distance in the plane of the sky}
\begin{figure}
    \includegraphics[trim= 0cm 0cm 0cm 0cm,width=\columnwidth]{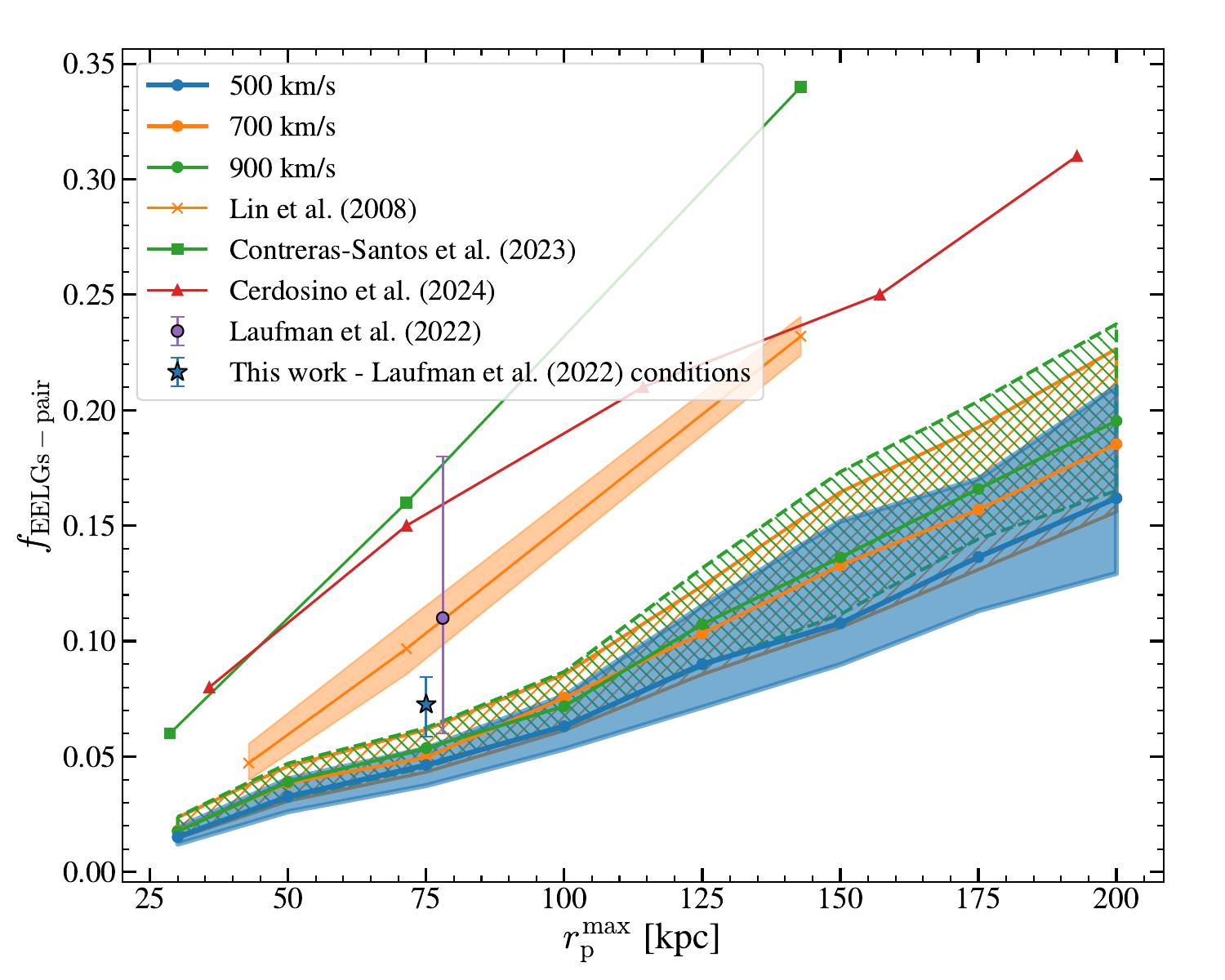}
    \caption{Pair fraction as a function of the maximum projected separation for this work and previous studies. The results of this work are shown in blue, orange, and green lines with dots for $\Delta v_\mathrm{max}$ of 500, 700 and 900 $\mathrm{km\,s^{-1}}$, respectively.  The shaded areas represent the $1\,\sigma$ error of the pair fraction obtained from bootstrap simulations. The orange line shows the empirical relation obtained by \cite{Lin2008} for DEEP2 galaxies, evaluated at the median redshift of the J-PAS sample of EELGs ($z_\mathrm{median}=0.41$). The orange shaded region shows the $1\,\sigma$ uncertainty of the relation. The pair fraction obtained by \cite{Contreras2022} for galaxies in the THREE HUNDRED simulations is shown in green. The red line shows the pair fraction obtained by \cite{Cerdosino2024} for the S-PLUS galaxy catalogue. The purple dot is the value obtained by \cite{Laufman2022} for a sample of GPs. As a comparison, the blue star is the result obtained for the J-PAS EELG sample restricted to the conditions applied by \citet[see the text for more details]{Laufman2022}.}
    \label{fig:pf_rp}
\end{figure}

In this subsection, we examine how the pair fraction of our EELG sample evolves with the maximum projected distance in the plane of the sky, that is, $r_p^{\rm max}$. We also analyse how our results relate to existing literature on this topic. Numerous studies have searched for companion galaxies in detail, establishing different criteria for selecting the samples of reference galaxies and companions, such as a given stellar mass range, a range for the mass ratio between the reference galaxy and its possible companions, a given limiting magnitude in a given photometric band, or, as in this study, specifying that the reference galaxy should be of a particular type, even for simulation-based studies \citep[][among many others]{Ellison2008,LopezSanJuan2010,LopezSanJuan2012,Rodriguez2020,Conselice2022,Gonzalez2023,Puskas2025}. 

Moreover, the criteria for choosing companion galaxies vary significantly across studies, particularly when contrasting merger-focused studies with those examining galaxy companions. In merger investigations, criteria such as maximum projected distances tend to be more restrictive ($r_p \sim 30-50\,\mathrm{kpc}$) than those used in studies that concentrate on galaxy companions ($r_p \gtrsim 100\,\mathrm{kpc}$). This discrepancy often complicates direct comparisons as the fraction of identified pairs is highly sensitive to the specific selection criteria employed.

These differences can be seen in Fig. \ref{fig:pf_rp}, where the pair fraction is shown against different values of the maximum projected separation obtained for the EELGs of J-PAS and for other studies. We present the redshift-dependent relationship derived by \cite{Lin2008} for DEEP2 Galaxy Redshift Survey \citep{Davis2003} galaxies with absolute magnitudes in the $B$-band ranging from -21 to -19 and a line of sight criterion of $\Delta v_\mathrm{max} = 500\, \mathrm{km\,s^{-1}}$. This relation was evaluated for the median redshift of the EELG sample ($z_\mathrm{median} = 0.41$). The work of \cite{Contreras2022}, who searched for galaxy pairs in the THREE HUNDRED simulations \citep{Cui2018}, is also included. To facilitate comparison, we use their results for $\Delta v_\mathrm{max} = 500\,\mathrm{km\,s^{-1}}$. Furthermore, we present the results of \cite{Cerdosino2024}, which are described in the previous section, who studied galaxies about two orders of magnitude more massive. 

In addition, the research conducted by \cite{Laufman2022} is also presented in Fig. \ref{fig:pf_rp}. This study examines a sample of 23 GPs from \cite{Cardamone2009} observed using the MUSE instrument, in the redshift range $0.15\lesssim z\lesssim0.33$. The authors aimed to determine the possible influence of companion galaxies within the MUSE field ($1' \times 1'$) centred on the EELGs. This limits  $r_p^\mathrm{max}$ between 134 \,kpc and 78 \,kpc over their redshift range, and they adopted  $\Delta v_\mathrm{max} = 300\,\mathrm{km\,s^{-1}}$. The results obtained for the J-PAS EELGs differ from those obtained by \cite{Laufman2022} for the MUSE GPs. The J-PAS EELGs tend to be more isolated than the MUSE GPs, even when the criteria imposed in the work of \cite{Laufman2022} are used (see the blue star in Fig. \ref{fig:pf_rp}). The pair fraction obtained for the J-PAS EELG sample, when $r_p^\mathrm{max}=78\,$kpc, $\Delta v_\mathrm{max} = 300\,\mathrm{km\,s^{-1}}$, and $0.15 < z < 0.33$, is $0.073^{+0.014}_{-0.012}$. 
Although this value is statistically consistent within the 1$\,\sigma$ uncertainties with that derived for the MUSE GPs sample ($0.11^{+0.07}_{-0.05}$), it systematically tends toward lower values.
This may essentially be due to the difference in the statistics of the objects analysed, since J-PAS has 235 EELGs that met the restrictions in \cite{Laufman2022}, significantly more than the 23 in the MUSE sample. Furthermore, we note that applying such a restrictive velocity threshold to photometric redshift data may exclude some physical pairs due to the limited redshift precision of J-PAS; thus, this comparison should be interpreted with caution.

Although direct comparisons cannot be made because of the different conditions of the samples studied, we still can gain some insight. It can be seen that the pair fraction of the J-PAS EELGs is smaller than that observed in general and more massive galaxies and also in simulations. This is in agreement with what is found in the literature, where more massive galaxies tend to show pair fractions four times larger than those of low-mass galaxies \citep{Rodriguez2015,Chamberlain2024}. 
This suggests that the low pair fraction of the EELGs may be, at least partly, a consequence of their low stellar masses. To test this, we constructed a control sample of J-PAS normal star-forming galaxies (SFGs) matched in stellar mass and redshift (see Appendix \ref{sec:appendixD}), for which we obtain $f_\mathrm{pair}^\mathrm{control}=0.26^{+0.13}_{-0.12}$, slightly higher than that of the EELGs but consistent within the uncertainties. However, the limited size of the control sample and its bias towards higher stellar masses prevent us from establishing whether EELGs are more isolated than other SFGs of similar mass. Our results are nevertheless consistent
with findings in the literature regarding both EELGs \citep{Amorin2015} and close analogues, such as local star-forming dwarf galaxies \citep[][among others]{Telles1995,Vilchez1995,Telles2000,Noeske2001,Koulouridis2013}.

Therefore, rather than implying that interactions are the dominant mechanism driving the EELG properties, 
our results indicate that a companion is not required for most EELGs. Interactions may act as a trigger in some systems, while in the rest the burst may be related to past or undetected interactions, gas accretion, or intrinsically bursty star formation.

\section{Analysis of the galaxy companions of EELGs}\label{sec:5}
\begin{figure*}
    \sidecaption
    \includegraphics[width=12cm]{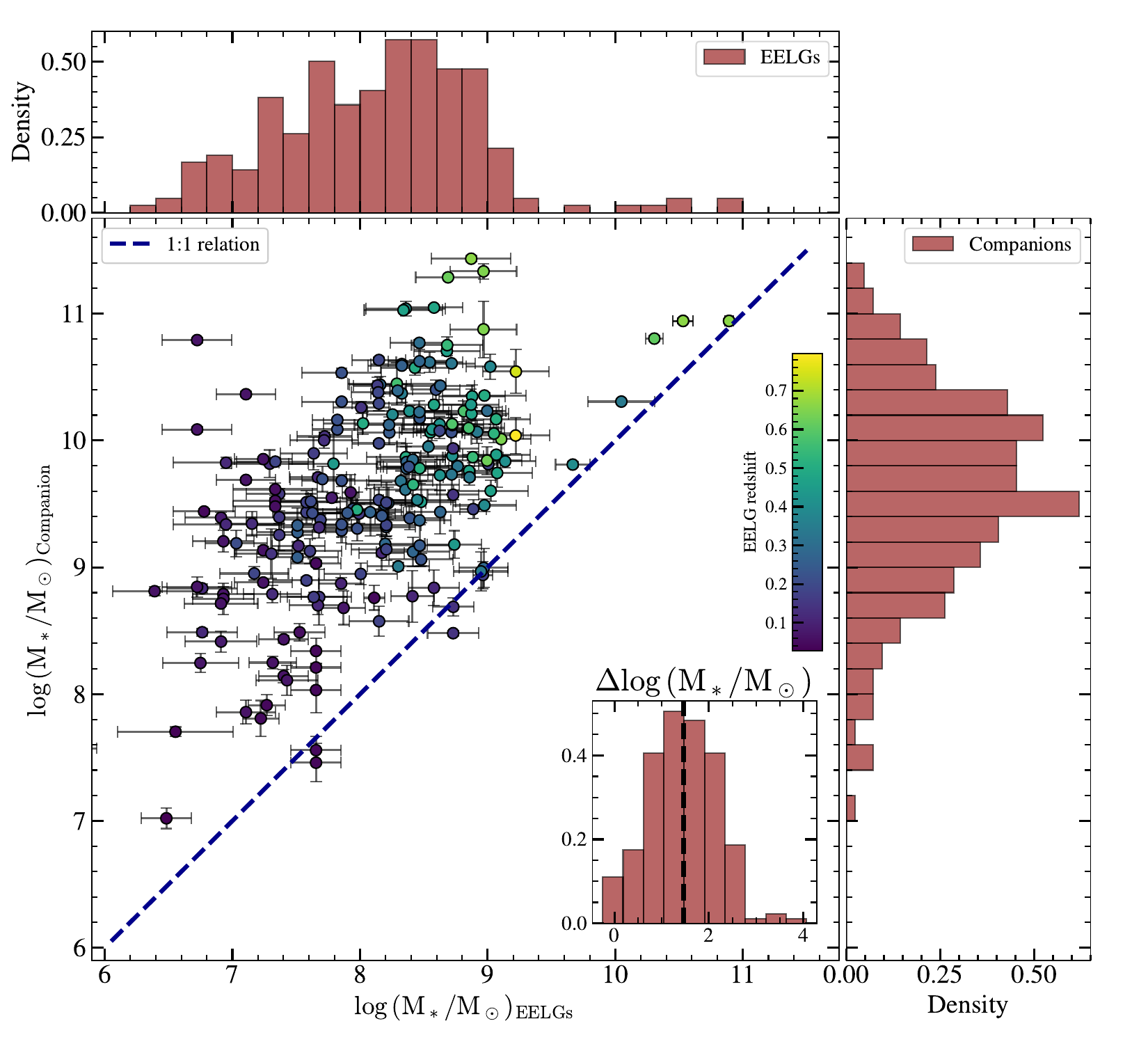}
    \caption{Comparison between the stellar masses derived for the selected EELGs and for their companions. In the central panel, the  dots are colour coded by the photometric redshift of the EELGs. The upper histogram shows the stellar mass distribution for the EELGs, and the vertical histogram corresponds to that of the companions. The inset shows the mass ratio between the companion galaxies and the EELGs, and the black dashed line marks the median of the distribution, corresponding to a value of $\Delta\log\left(\mathrm{M}_*/\mathrm{M}_\odot\right)\sim1.5$.}
    \label{fig:mass_comparison}
\end{figure*}

In this section, we focus on the EELGs with $f_\mathrm{pair} > 0$ after correcting for the random contribution (see Sect.~\ref{sec:4.1}). This subsample comprises 159 EELGs, with a median $f_\mathrm{pair}$ of 0.21 and a 75th percentile of 0.47. We compare their stellar masses, SFRs, sSFRs, dust attenuation, and 
SED-based
metallicities with those of their companion galaxies, 
using the homogeneous SED fits described in Sect.~\ref{sec:comp_sel}.
This analysis allows us to characterise the physical nature of the companion population and to assess whether they represent a distinct galaxy type or a diverse set of systems.

\subsection{Stellar mass comparison}
First, we studied the stellar masses of the galaxies derived from the SED fittings with \texttt{CIGALE}. Figure \ref{fig:mass_comparison} shows the stellar masses obtained for both EELGs and companion galaxies, coloured by the redshift of the EELGs. As can be seen in the central panel, companion galaxies tend to be systematically more massive than EELGs, with median stellar masses of $\log\left(\mathrm{M}_*/\mathrm{M}_\odot\right)=8.32^{+0.57}_{-0.94}$ for the EELGs and  $\log\left(\mathrm{M}_*/\mathrm{M}_\odot\right)=9.61^{+0.73}_{-0.78}$ for the companion galaxies, where uncertainties correspond to the 16th–84th percentiles of the distributions. The observed mass ratio between EELGs and their companions is also notable (inset histogram in Fig. \ref{fig:mass_comparison}), with a median  $\Delta\log\left(\mathrm{M}_*/\mathrm{M}_\odot\right)=1.47$, that is, a difference of more than one order of magnitude between companions and EELGs. 

This differs from the result of \cite{Gupta2023}, who found that, in a sample of 19 EELGs at high redshift ($2.5 < z < 4$), the EELGs tend to be more massive than their companions, with an average mass ratio between the most massive companion and the EELG of $\sim0.65$. This discrepancy may be due to the selection criteria the authors used for the companion search, where they used $r_p^\mathrm{max} < 40\,\mathrm{kpc}$ and $\Delta v_{\rm max} = 10\,000\,\mathrm{km\,s^{-1}}$. 
In addition, the two studies probe very different cosmic epochs. While the J-PAS sample lies at $z \sim 0.4$, \cite{Gupta2023} explore galaxies in the early Universe, when galaxy populations were generally more gas-rich and less evolved. Therefore, part of the discrepancy could also be related to underlying evolutionary differences, rather than arising solely from methodological effects.
Additionally, recent results by \cite{ArroyoPolonio2026} based on MUSE observations of a sample of 24 GPs, in a redshift range of $0.15 < z < 0.33$, report the detection of 11 companions within $\Delta v_\mathrm{max} = 500\,\mathrm{km\,s^{-1}}$, showing a companion fraction of $0.33^{+0.11}_{-0.08}$. In that study, nearly 50\% of the companion galaxies are found to have stellar masses lower than those of the GPs. This apparent difference with respect to our J-PAS results can be naturally explained by the small field of view and high surface-brightness sensitivity of MUSE, which favour the detection of faint, low-mass companions in the immediate vicinity of the target galaxies.

Although the overall pair fraction in our sample is low ($f_\mathrm{EELGs\text{-}pair}=0.162^{+0.048}_{-0.033}$ for $r_p^\mathrm{max} = 200\,\mathrm{kpc}$ and $\Delta v_{\rm max} = 500\,\mathrm{km\,s^{-1}}$), indicating that most EELGs do not have a close companion, the systems that contribute to this fraction consistently show companions that are more massive. This suggests that while interactions are not common among EELGs, when they occur they often involve a low-mass EELG in the vicinity of a more massive galaxy. These low-mass--high-mass galaxy pairings could still play a role in triggering or enhancing the current starburst in a subset of the population.

It is also important to note that our companion sample is restricted to galaxies within the stellar-mass completeness limit of the J-PAS survey. This constraint naturally biases the analysis against detecting lower-mass companions. All sources in the EELG sample are selected from the J-PAS parent catalogue, which is constructed in dual mode using the broad $i_{\rm SDSS}$ band for detection, and are therefore subject to the same initial broadband detection limits as the rest of the galaxy population. However, the large width of the $i_{\rm SDSS}$ filter ($\sim1000\,\mathrm{\AA}$), combined with the extremely strong emission lines characteristic of EELGs, can significantly boost the broadband flux when a bright optical line falls within the filter. As a result, galaxies with intrinsically faint continua but very strong emission lines are more likely to be detected than galaxies of similar mass and redshift lacking such lines. This asymmetry in detectability could enhance the apparent mass difference between EELGs and their companions and could also make pairs with smaller companions appear as isolated EELGs. Overall, the observed mass relation indicates that 
the detected companions are systematically more massive, which is partly expected from the completeness limit imposed on the companion sample.

\subsection{SFR, sSFR, dust attenuation, and metallicity comparison}
\begin{figure*}
    \sidecaption
    \includegraphics[width=12cm]{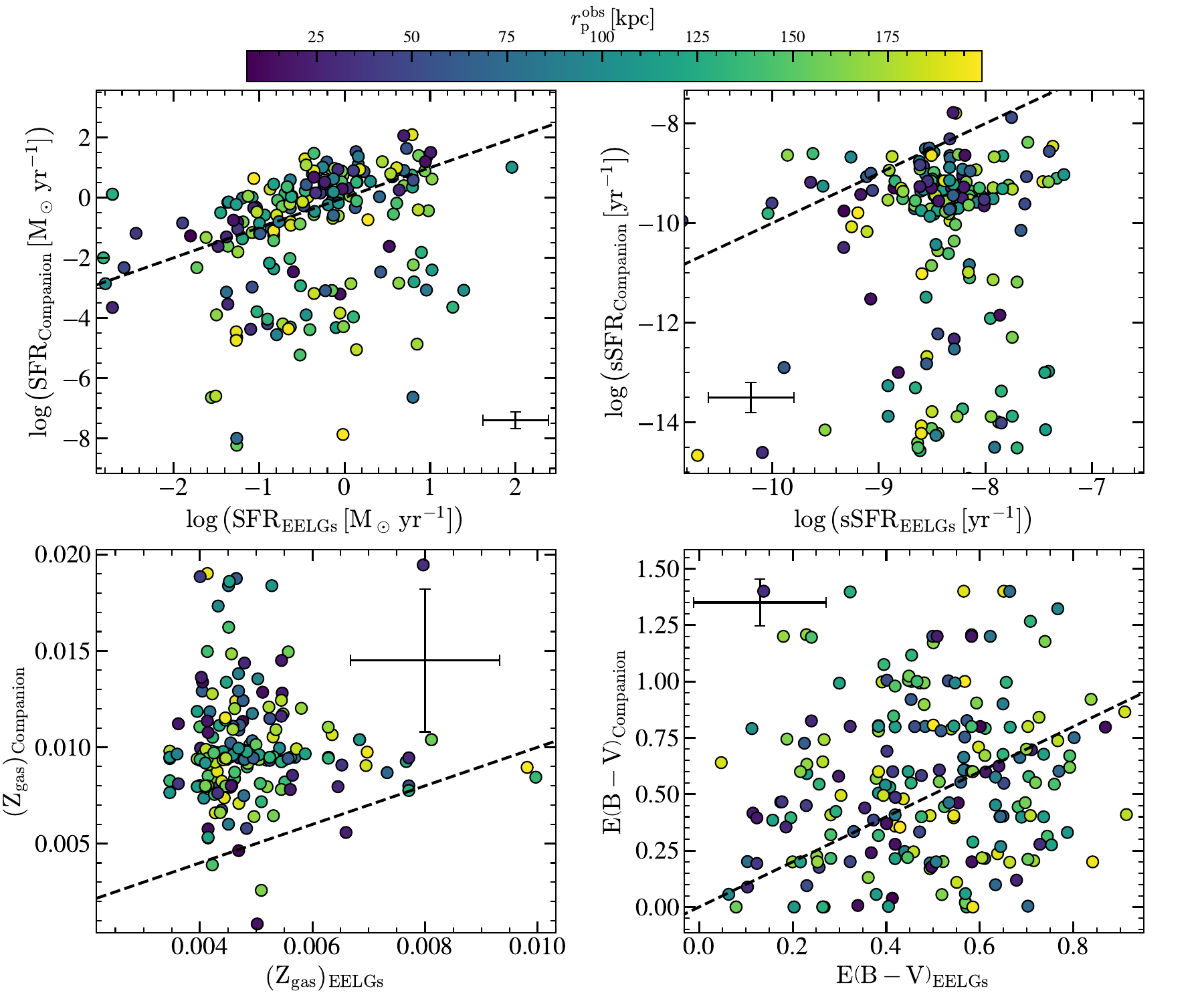}
    \caption{Comparison of SFR (top-left), sSFR (top-right), \(Z_{\mathrm{gas}}\) (bottom-left), and E(B--V) (bottom-right) between EELGs and their companion galaxies. Points are colour-coded by the observed projected separation, and error bars represent the median uncertainties derived from the SED fittings with \texttt{CIGALE}. The black dashed lines represent the 1:1 relation.}
    \label{fig:properties_comparison}
\end{figure*}

In this section, we compare some physical properties of EELGs and their companions, focusing on their SFR, sSFR, dust attenuation [E(B--V)], and gas-phase metallicities (\(Z_{\mathrm{gas}}\)). The purpose is to determine whether companion galaxies tend to share similar physical characteristics, for example, if the companions of EELGs are typically also strong star formers with low dust attenuation and metal-poor systems or not. 

Figure \ref{fig:properties_comparison} compares these quantities for both members of each pair. The upper panels display the relations for SFR and sSFR, while the lower panels correspond to \(Z_{\mathrm{gas}}\) and E(B--V). The dashed line marks the 1:1 relation, and the points are colour-coded according to the projected separation between the galaxies. The general picture is that, although EELGs and their companions can sustain similar levels of star formation in absolute terms, their sSFRs differ markedly. The SFR panel reveals that, within the scatter, $\sim65\,\%$ of EELGs and their companions form stars at comparable absolute rates. However, the sSFR panel shows a clear systematic offset, with $\sim92\,\%$ of the companion galaxies lying below the 1:1 relation. This indicates that for a given pair, the companion preferentially has a lower sSFR than the EELG. Since sSFR scales inversely with stellar mass, this behaviour is consistent with the EELG being the less massive but more actively star-forming member of the pair.

The inclusion of metallicity provides an additional dimension to this comparison. However, it should be noted that the adopted gas-phase metallicity \(Z_{\mathrm{gas}}\) values are proxies for the chemical enrichment of galaxies, as they are derived from SED fitting rather than direct spectroscopic diagnostics, and that they are sampled on a coarse grid of six values (Table \ref{tab:CIGALE}), so small differences should not be over-interpreted.
Within this framework, the metallicity panel shows that companion galaxies generally exhibit higher inferred $Z_{\mathrm{gas}}$ values than the EELGs.
In particular, 96\,\% of the pairs lie above the 1:1 relation, with a median difference of $\Delta Z_\mathrm{gas} = 0.0046\pm0.0007$ and $P(\Delta Z_\mathrm{gas}\leq0)<10^{-4}$ from Monte Carlo resampling of the uncertainties and bootstrapping. However, the difference is individually significant at more than $2\sigma$ in only $\sim4\%$ of the pairs, and for most companions the uncertainty of $Z_{\mathrm{gas}}$ is comparable to the dispersion of the prior in the \texttt{CIGALE} input parameters, indicating that their metallicity is poorly constrained by the photometry and their estimate is driven towards the mean of the grid. Restricting the comparison to the 31 companions with well constrained metallicities, the offset persists. For 30 of 31 pairs, the result lies above the 1:1 relation, with $\Delta Z_\mathrm{gas} = 0.0041\pm0.0008$.
This tentative offset is consistent with the fact that EELGs are chemically younger systems that are still in the process of enriching their ISM through ongoing star formation. Conversely, their companions, often more massive and with lower sSFR, tend to be more chemically enriched, as expected from longer evolutionary timescales and more efficient past star formation.

In contrast, the dust attenuation of EELGs and their companions is comparable. The companions show higher E(B--V) values in only $\sim$53\% of the pairs, with a median difference of $\Delta$E(B-V)$=0.03$. We note that the E(B--V) values derived for the EELGs are sensitive to the attenuation prescription adopted in the SED fitting (Appendix~\ref{sec:appendixA}), which highlights the importance of fitting both members of the pairs homogeneously.

Inspection of Fig. \ref{fig:properties_comparison} indicates that there is no strong dependence on projected separation. This absence of correlation suggests that the differences in SFR, sSFR, \(Z_{\mathrm{gas}}\), and E(B--V) 
do not depend on the current projected separation between the EELG and its detected companion.
The overall similarities in SFR 
and dust attenuation 
between both components, together with the systematic differences in sSFR  
and the tentative offset in metallicity estimates,
are consistent with an evolutionary connection in which EELGs represent transient, high-efficiency episodes within low-mass galaxies that coexist with (but are not necessarily driven by) their more evolved companions.

\subsection{Do companion galaxies affect the emission-line strength of EELGs?}\label{sec:5_3}
\begin{figure*}
    \includegraphics[trim= 0cm 0cm 0cm 0cm,width=\textwidth]{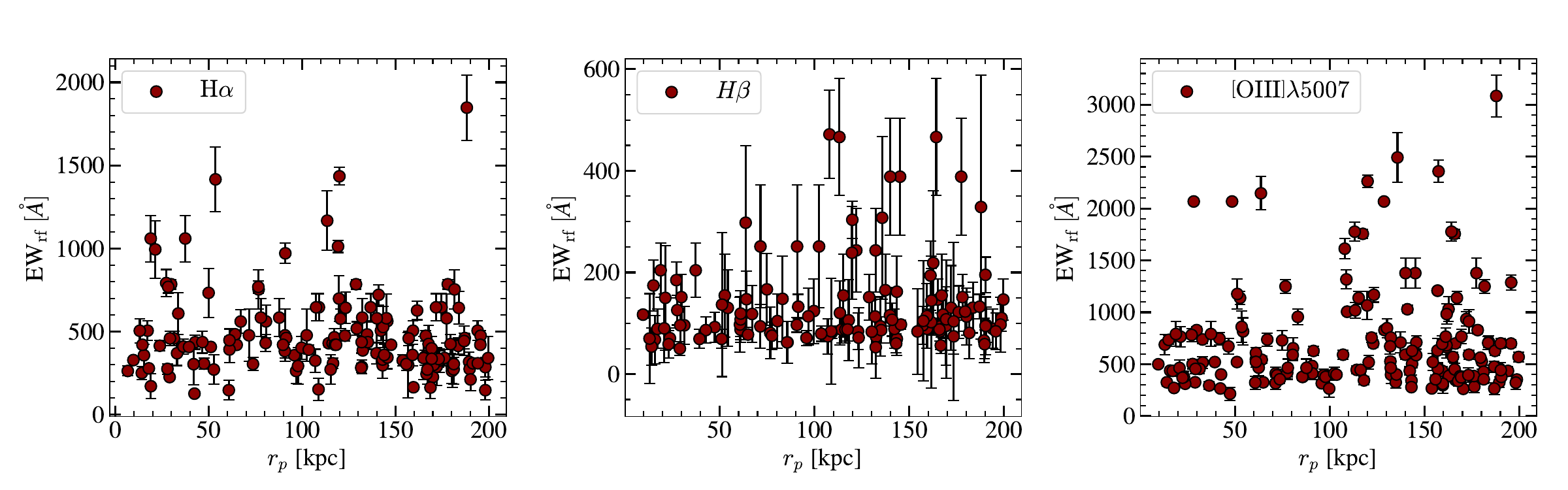}
    \caption{Rest-frame EWs of H$\alpha$, H$\beta$, and [\ion{O}{III}]\,$\lambda5007$ as a function of the projected separation (\(r_p\)) between each EELG and its companions. The panels contain different subsamples (106, 97, and 133 EELGs, respectively), since H$\alpha$ is only covered by J-PAS up to $z\sim0.4$ (see Sect. \ref{sec:5_3} for details).}
    \label{fig:EW_analysis}
\end{figure*}

We explored whether the strength of the emission lines in the EELGs depends on the distance to a nearby companion.
To do this, we used the rest-frame EWs obtained by \cite{AntonioG}, where they were derived from the J-PAS photometry by subtracting the continuum predicted by the \texttt{CIGALE} models from the flux of the filter containing each line, and correcting for the contribution of neighbouring lines (see their Sects. 4.4 and 4.5 for more details).

Figure \ref{fig:EW_analysis} shows the rest-frame EWs of H$\alpha$, H$\beta$, and [\ion{O}{III}]\,$\lambda5007$ of the EELGs as a function of the projected separation to their companions. These lines trace the ongoing star formation and the ionisation state of the gas, with higher EWs typically indicating galaxies undergoing intense, recent bursts of star formation. 
The three panels do not contain the same EELG sample, since the availability of each line depends on the J-PAS spectral coverage. In particular, H$\alpha$ falls within the J-PAS wavelength range only up to $z\sim0.4$. The H$\alpha$ panel includes 106 EELGs (150 pairs) at $0.03\leq z \leq0.39$ (median $z=0.19$), the H$\beta$ panel 97 EELGs (128 pairs) at $0.03\leq z \leq 0.80$ (median $z=0.29$), and the [\ion{O}{III}]\,$\lambda 5007$ panel 133 EELGs (172 pairs) at $0.03 \leq z \leq 0.80$ (median $z=0.28$). All EELGs in the H$\beta$ panel are also included in the [\ion{O}{III}] panel, and 59 EELGs are common to all three panels.

No clear monotonic relation is found between the EW and the current projected separation for any of the three lines. Using the closest companion of each EELG, we obtain Spearman correlation coefficients of $\rho_S = -0.14\;(p=0.14),+0.09\;(p=0.36)$, and $0.00\;(p=0.98)$ for H$\alpha$, H$\beta$, and $[\ion{O}{III}]\,\lambda5007$, respectively. These results also hold when all pairs are considered. To verify that the comparison between lines is not affected by the different subsamples, we repeated the analysis for the 59 EELGs common to the three lines and for the 97 EELGs common to H$\beta$ and [\ion{O}{III}]\,$\lambda5007$. In both cases, no significant correlation was found for any line ($|\rho_S|\leq 0.17$, $p\geq 0.20$).
This lack of trend suggests that, within the limits of the J-PAS data set, the emission-line strength in EELGs 
does not depend on the current projected distance to a detected companion. 

Although some of the most extreme EWs appear at small separations, particularly in H$\alpha$, similarly high values are also found at the largest separations, especially for H$\beta$ and more prominently for [\ion{O}{III}]\,$\lambda5007$. These objects are particularly intriguing, as they may trace diverse physical scenarios, ranging from isolated, metal-poor starbursts to systems experiencing transient post-interaction phases. The forthcoming J-PAS data releases, covering substantially larger areas, will allow us to determine whether this population becomes statistically significant or remains a minority tail in the EW distribution. 

In general, the presence and distance of a companion do not appear to produce any systematic enhancement in line strength, indicating that 
the presence of a detectable companion is not a necessary condition for the extreme emission-line phase.
This result is consistent with published studies showing that although close pairs can exhibit modest enhancements in H$\alpha$ or [\ion{O}{III}]\,$\lambda5007$ emission \citep{Lambas2003, Woods2006, Ellison2008, Purkayastha2022}, the extreme EWs observed in EELGs have been proposed to be primarily linked to their compactness, low metallicity, and high star-formation efficiency rather than to environmental effects \citep{Amorin2015, Laufman2022}. 
In this sense, the data suggest that, while the presence of a companion may influence the long-term evolution of some EELGs, the current projected proximity does not modulate their ongoing star formation or current ionisation state.

\section{Conclusion} \label{sec:6}
In this work, we have presented the first statistical characterisation of the local environments of EELGs using data from the J-PAS survey. We used a sample of 917 EELGs identified by \cite{AntonioG}. Exploiting the unique capabilities of J-PAS—its contiguous narrow-band filter system and precise photometric redshifts—we developed a probabilistic framework based on the PPF to assess the likelihood that each EELG has a physically associated companion. This methodology makes full use of the individual PDF of the photometric redshift, allowing for a statistically robust determination of pair fractions while accounting for projection effects, completeness corrections, and photometric uncertainties.

After applying completeness and quality masks and correcting for random projections through simulations, we derived a corrected pair fraction of $f_{\mathrm{EELGs\text{-}pair}} = 0.162^{+0.048}_{-0.033}$ for companions within projected separations $r_p \leq 200$\,kpc and line-of-sight velocity differences $\Delta v_\mathrm{max} = 500$\,km\,s$^{-1}$. 
The pair fraction indicates that only about 16\% of the sample has detectable companions within the adopted separation, velocity, and completeness limits. This result should be interpreted as a lower limit, as companions below the J-PAS stellar-mass completeness limit may be missed more easily given that the strong emission lines of EELGs boost their detectability relative to galaxies of similar mass. Within these limits, the low incidence of detectable companions suggests that interactions with massive neighbours are not the dominant trigger of the EELG phase for most of the sample.
This result is consistent with recent environmental studies of emission-line galaxies in other surveys, which also find that enhanced star formation activity is not systematically associated with mergers or close interactions \citep{Kimsey2026}.

When EELGs are found to possess nearby companions, these are, on average, more massive and 
tend to show higher SED-based metallicity estimates
within the limits of our J-PAS-based companion selection. 
However, this result is strongly affected by the stellar-mass completeness of the companion sample, since only companions more massive than the typical EELGs can be detected in J-PAS, and fainter, lower-mass companions may remain undetected, as suggested by recent works revealing low-mass satellites around EELGs \citep{Gupta2023,ArroyoPolonio2026}. Bearing this limitation in mind, the observed asymmetry suggests that EELGs typically represent the lower-mass members of the detected pairs, potentially acting as satellites or recently accreted components within larger haloes.
Despite the potential for tidal interactions or gas exchange in such configurations, we find no significant correlation between the emission-line EW and the current projected separation of the detected pairs 
(Spearman $|\rho_S|\leq0.14$, $p\geq0.14$). 
This absence of a trend suggests that even when companions exist, the intensity of the ongoing starburst 
is not modulated by the projected distance to the detected companions.
Nevertheless, we cannot rule out the possibility that the EELG phase may preferentially occur during late-stage mergers, coalescence, or post-merger phases of gas-rich mergers, where the original interacting configuration is no longer identifiable as a separate pair. As a result, our conclusions are necessarily statistical in nature, and disentangling causal connections between interactions and bursty star formation will require deep, spatially resolved imaging and spectroscopic kinematic studies.

Taken together, and in a statistical sense, our results are consistent with a picture in which EELGs represent a transient yet recurrent evolutionary stage of low-mass galaxies, characterised by rapid, stochastic episodes of massive gas accretion and feedback from the halo. These processes can maintain very high ionisation rates and low metallicities without requiring external triggers \citep[e.g.][]{Ceverino2016}. This behaviour appears consistent with theoretical models of dwarf galaxy evolution \citep[e.g.][]{LLerena2023} and with recent observations of star-forming galaxies at $z > 6$ \citep[e.g.][]{Schaerer2022}, reinforcing the interpretation of EELGs as nearby analogues of early-universe systems that contributed to cosmic reionisation. 

This work also demonstrates the potential of J-PAS for studies of galaxy environment and evolution using purely photometric data. The combination of contiguous narrow-band coverage, accurate photometric redshifts, and large survey area enables the construction of statistically significant samples of rare populations such as EELGs and the quantitative characterisation of their surroundings. The probabilistic methodology developed here can be applied to other galaxy types and extended to future J-PAS data releases, which will cover substantially larger areas.

\begin{acknowledgements}

We thank the anonymous referee for their useful report. 
MGO, AGA, RGD, GMS, L.A.D.G., AAC, and IM acknowledge financial support from the State Agency for Research of the Spanish MCIU through Center of Excellence Severo Ochoa award to the Instituto de Astrofísica de Andalucía CEX2021-001131-S funded by MCIN/AEI/10.13039/501100011033.
MGO and AGA acknowledge support from the research grant PID2022-136598NB-C32 ("Estallidos8").
MGO also acknowledges the support by the project ref. AST22\_00001\_Subp\_11 funded from the EU – NextGenerationEU, PPCC Junta de Andalucía. 
JAFO acknowledges financial support by the Spanish Ministry of Science and Innovation (MCIN/AEI/10.13039/501100011033), by ``ERDF A way of making Europe'' and by ``European Union NextGenerationEU/PRTR'' through the grants PID2021-124918NB-C44 and CNS2023-145339; MCIN and the European Union -- NextGenerationEU through the Recovery and Resilience Facility project ICTS-MRR-2021-03-CEFCA.
RGD, GMS, and L.A.D.G. acknowledge the financial support from the project PID2022-141755NB-I00.
FJSR acknowledges financial support by PREP2023-001336 funded by MICIU/AEI/10.13039/501100011033 and ESF+, associated to Grant PID2023-147386NB-I00 funded by MICIU/AEI/10.13039/501100011033 and by ERDF/EU.
ACS acknowledges support from FAPERGS (grants 23/2551-0001832-2 and 24/2551-0001548-5), CNPq (grants 314301/2021-6, 312940/2025-4, 445231/2024-6, and 404233/2024-4), and CAPES (grant 88887.004427/2024-00).
IM acknowledges financial support from the project PID2022-140871NB-C21.
SZ acknowledges the financial support provided by the Governments of Spain and Aragón through their general budgets and the Fondo de Inversiones de Teruel, the Aragonese Government through the Research Group E16\_23R, and the Spanish Ministry of Science and Innovation and the European Union - NextGenerationEU through the Recovery and Resilience Facility project ICTS-MRR-2021-03-CEFCA.
AAC acknowledges financial support from the Severo Ochoa grant CEX2021-001131-S and the project PID2023-153123NB-I00 funded by MCIN/AEI/10.13039/501100011033 and the FSE+.
This paper has gone through internal review by the J-PAS collaboration. Based on observations made with the JST/T250 telescope and JPCam at the Observatorio Astrofísico de Javalambre (OAJ), in Teruel, owned, managed, and operated by the Centro de Estudios de Física del Cosmos de Aragón (CEFCA). We acknowledge the OAJ Data Processing and Archiving Unit (UPAD) for reducing and calibrating the OAJ data used in this work. Funding for the J-PAS Project has been provided by the Governments of Spain and Arag\'on through the Fondo de Inversiones de Teruel; the Aragonese Government through the Research Groups E96, E103, E16\_17R, E16\_20R, and E16\_23R; the Spanish Ministry of Science and Innovation (MCIN/AEI/10.13039/501100011033 y FEDER, Una manera de hacer Europa) with grants PID2021-124918NB-C41, PID2021-124918NB-C42, PID2021-124918NA-C43, and PID2021-124918NB-C44; the Spanish Ministry of Science, Innovation and Universities (MCIU/AEI/FEDER, UE) with grants PGC2018-097585-B-C21 and PGC2018-097585-B-C22; the Spanish Ministry of Economy and Competitiveness (MINECO) under AYA2015-66211-C2-1-P, AYA2015-66211-C2-2, and AYA2012-30789; and European FEDER funding (FCDD10-4E-867, FCDD13-4E-2685).
\end{acknowledgements}

\bibliographystyle{aa}
\bibliography{biblio} 

\newpage
\begin{appendix} 
\section{\texttt{CIGALE} input parameters}\label{sec:appendixA}
\begin{table*}
    \centering
    \caption{Parameters and models used in the SED fitting with \texttt{CIGALE} for the EELGs with companions and their companion candidates.}
    \label{tab:CIGALE}
    \begin{tabular}{c c c}
    \hline \hline
    Model used & Parameter information & Values\\
    \hline
    SFH:                        & e-folding time main population  & 5, 10, 20, 50, 100, 1000, 4000, 8000 [Myrs] \\
    delayed SFH with            & age of the main population      & 30, 100, 200, 1000, 4000, 8000 [Myrs]\\
    optional exponential burst  & e-folding time of late burst    & 0.5, 1, 25, 50, 100, 1000     [Myrs] \\
                                & age of the late burst           & 1, 3, 6, 10, 50 [Myrs] \\
                                & mass fraction late burst        & 0.0, 0.01, 0.1, 0.15, 0.25, 0.5   \\
    \hline
    SSP:                       &    IMF          &   \cite{Chabrier2003}                                   \\
    \cite{Bruzual2003}  &    metallicity  &  0.0001, 0.004, 0.008, 0.02, 0.05    \\
    \hline
    Nebular component & ionisation parameter $\left(\log U\right)$ & -4, -3, -2, -1.5  \\
                    & gas metallicity $\left(Z_\mathrm{gas}\right)$ & 0.0001, 0.001, 0.004, 0.008, 0.014, 0.02 \\
                    & electron density $\left(n_e\right)$ & 100\\
    \hline
    Dust attenuation:    &  E(B-V) lines  & 0, 0.1, 0.2, 0.3, ..., 0.9, 1.0 \\
    power law modified   &  E(B-V) factor (line to continuum)  & 0.44 \\
     \cite{Calzetti2000} &  UV bump wavelength                  & 217.5 [nm] \\
                                 &  UV bump FWHM                       & 35 \\
                                 &  UV bump amplitude  & 0, 1.5, 3 \\
                                 &  Power law modification slope  & -0.2, 0 \\
                                 &  $R_{\rm V}$  & 3.1 \\
    \hline
    Dust emission &         AGN fraction & 0, 0.1, 0.3, 0.5, 0.8, 0.9\\
    \cite{Dale2014} &       $\alpha$ slope (dust heating) & 0.125, 1.0, 2.0 \\
    \hline

    \end{tabular}
\end{table*}

We compared the properties of the 159 EELGs with companions derived with this configuration with those of \cite{AntonioG}, which were obtained with a configuration specifically tailored to EELGs. The median differences (new minus original) and normalised median absolute deviations are $\Delta\log \mathrm{M_*} = -0.10\,(0.29)$ dex, $\Delta \log \mathrm{SFR}=+0.09\,(0.25)$ dex, $\Delta \mathrm{Z_{gas}}=+0.0001\,(0.0006)$, and $\Delta \mathrm{E(B-V)}=+0.27\,(0.16)$. The stellar masses, SFRs, and metallicity estimates are therefore consistent within the uncertainties, whereas E(B--V) shows a systematic offset that arises from the different attenuation prescriptions adopted in both works. For this reason, all comparisons between EELGs and companions in Sect.~\ref{sec:5} are based on homogeneous fits.

\section{\texttt{Odds} sampling rate (OSR) function for the J-PAS survey} \label{sec:appendixB}
\begin{figure}[h!]
    \centering
    \includegraphics[trim= 0cm 0cm 0cm 0cm,width=\columnwidth]{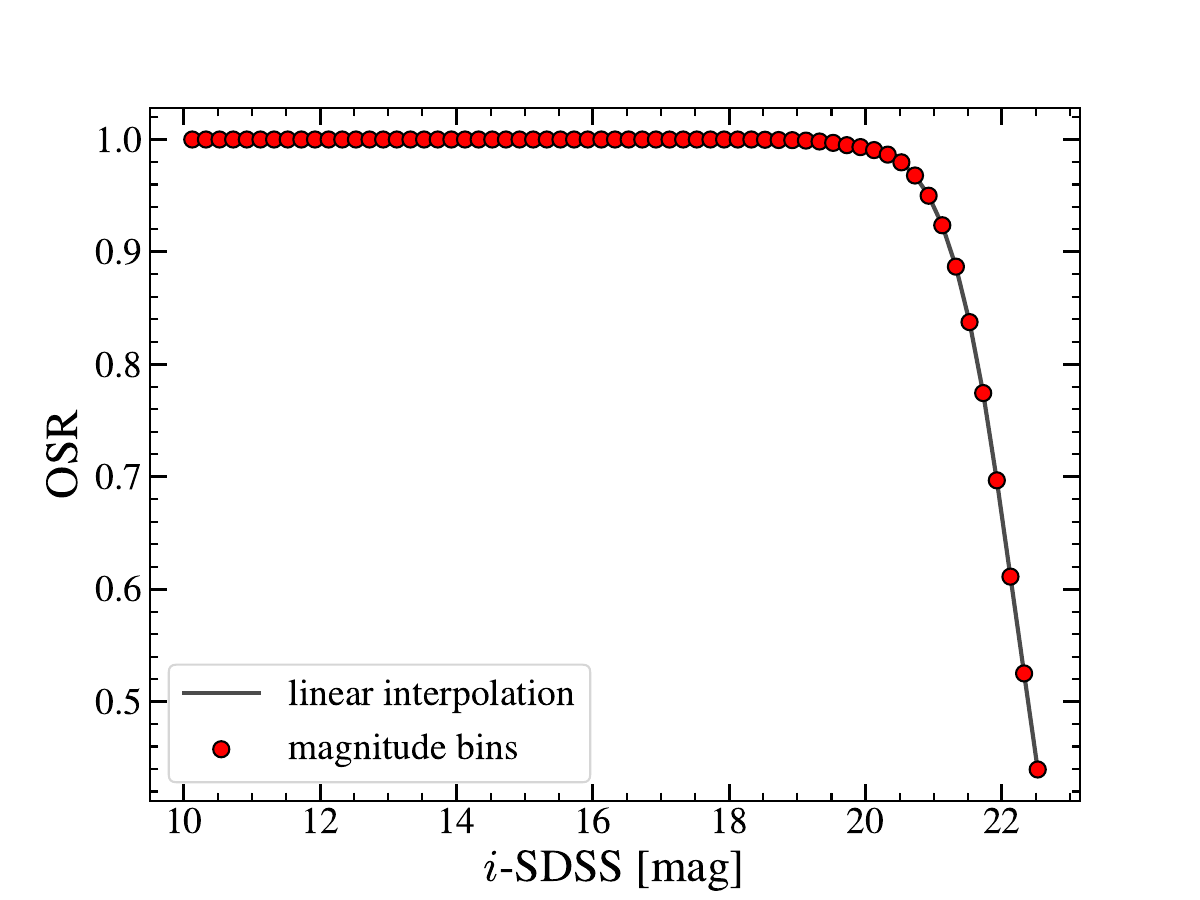}
    \caption{The OSR function obtained for the J-PAS survey, where the $i$-SDSS magnitude is related with the error in the photometric redshift. All objects in the J-PAS catalogue with calculated photometric redshift were used. To obtain the function, we performed an $i$-SDSS band magnitude binning, $\Delta m=0.2$ (red dots), and a linear interpolation to the points (black line).}
    \label{fig:OSR}
\end{figure}

\section{Spectroscopic validation of the probabilistic pair selection} \label{sec:appendixC}
\begin{figure}[h]
    \centering
    \includegraphics[trim= 0cm 0cm 0cm 0cm,width=\columnwidth]{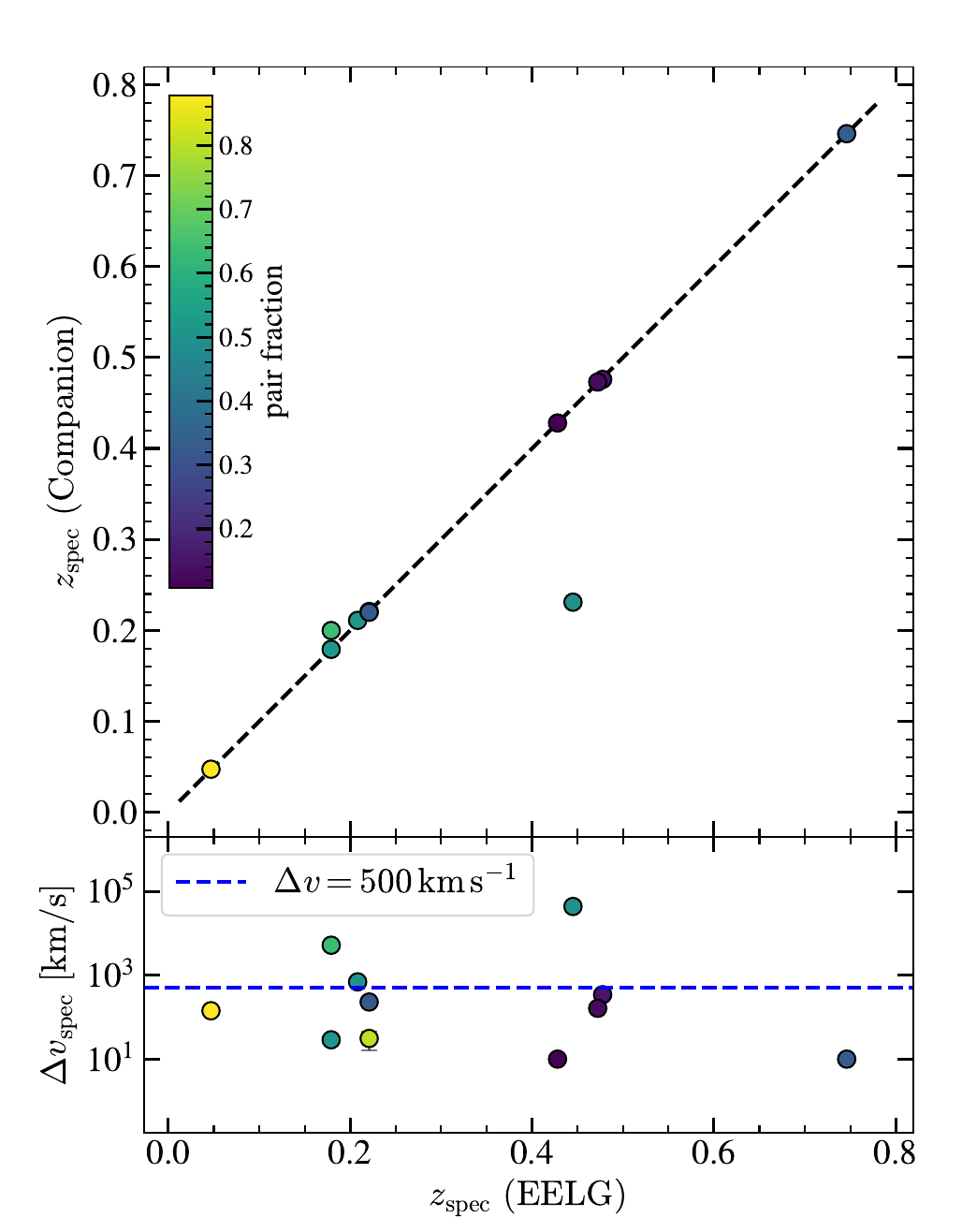}
    \caption{(Top) Comparison between the spectroscopic redshifts of the EELGs and those of their companion candidates with $f_\mathrm{pair}>0$. Most systems lie close to the one-to-one relation (black dashed line), indicating that photometric and spectroscopic selections are consistent, with only one outlier showing a large discrepancy. (Bottom) Line-of-sight spectroscopic velocity difference, $\Delta v_{\rm spec}$, as a function of the EELG spectroscopic redshift. The horizontal blue dashed line marks the adopted threshold of $\Delta v_{\rm spec}=500$\,km\,s$^{-1}$ used to define physical pairs.  Eight out of eleven pairs (success fraction $0.73^{+0.13}_{-0.19}$) fall below this limit.}
    \label{fig:spec_check}
\end{figure}
    
A limited but informative validation of the probabilistic pair selection was performed using the subset of EELGs and companion candidates for which $z_{\rm spec}$ is available in public catalogues. For each pair where both members have $z_{\rm spec}$ and $f_\mathrm{pair}>0$, we compute the line-of-sight velocity difference as:
\begin{equation}
\Delta v_{\rm spec}=c\times\frac{|z_{\rm spec,\,comp}-z_{\rm spec,\,EELG}|}{1+z_{\rm spec,\,EELG}},
\end{equation}
where $c$ is the speed of light. This comparison allows us to assess the performance of the PPF by verifying whether pairs identified using photometric redshifts also satisfy the spectroscopic criteria for physical association.

Figure~\ref{fig:spec_check} shows the 11 EELG–companion pairs with $f_\mathrm{pair}>0$ for which both galaxies have available spectroscopic redshifts in the literature. The top panel compares the spectroscopic redshifts of EELGs and their companions, revealing that most systems lie close to the one-to-one relation, with only one pair showing a significant discrepancy. The bottom panel displays the distribution of $\Delta v_{\rm spec}$ as a function of the EELG redshift. The adopted criterion for $\Delta v_{\rm spec} < 500$\,km\,s$^{-1}$ is indicated by the horizontal blue dashed line. 
Eight out of the eleven pairs fall below this threshold, corresponding to a success fraction of $0.73^{+0.13}_{-0.19}$ (68\% binomial confidence interval).

Despite the small statistics, these results are encouraging. The close agreement between the spectroscopic and photometric redshifts for most pairs supports the reliability of the probabilistic methodology and indicates that the PPF formalism is able to recover physically associated companions even in the presence of photometric redshift uncertainties. Although the limited number of pairs does not allow us to rule out biases in the identification of physical companions, this validation provides additional confidence in the environmental trends derived from the J-PAS EELG sample.

\section{Control sample of J-PAS galaxies}\label{sec:appendixD}
\begin{figure*}
    \centering
    \includegraphics[width=\columnwidth]{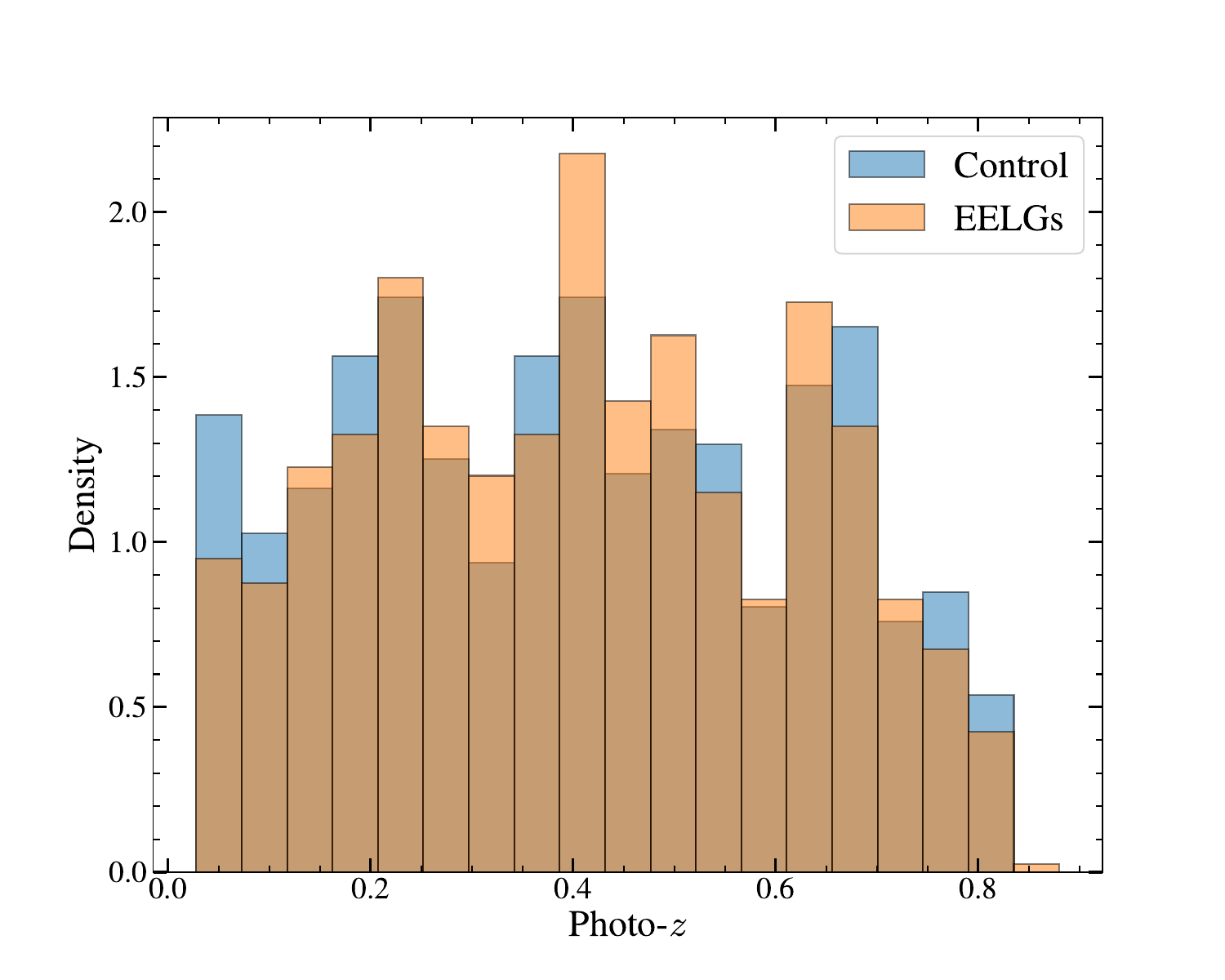}
    \includegraphics[width=\columnwidth]{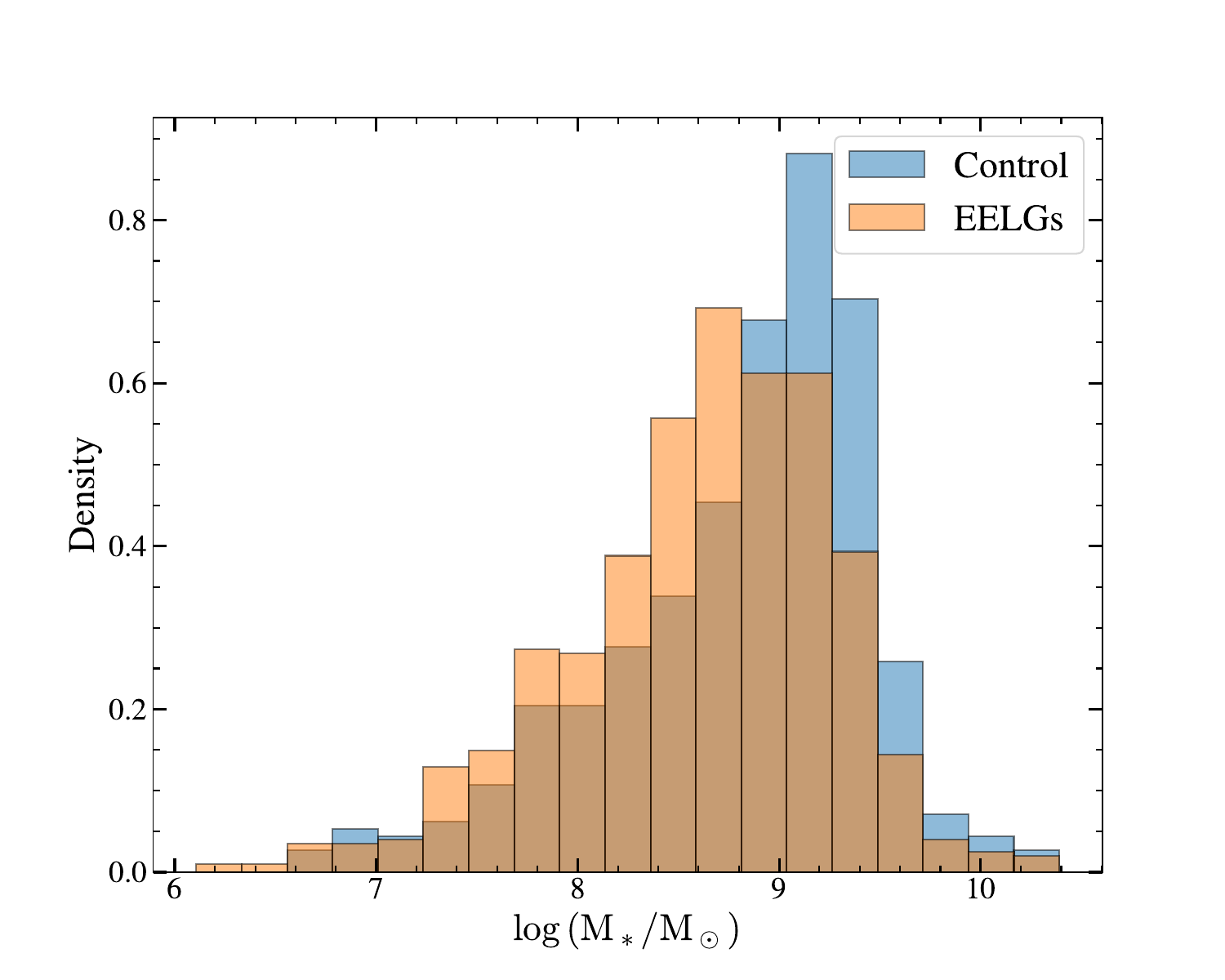}
    \caption{Distributions of the photometric redshift (left) and stellar mass (right) of the control sample (blue) and the EELG sample (orange).}
    \label{fig:control_distributions}
\end{figure*}

\begin{figure}
    \centering
    \includegraphics[width=\columnwidth]{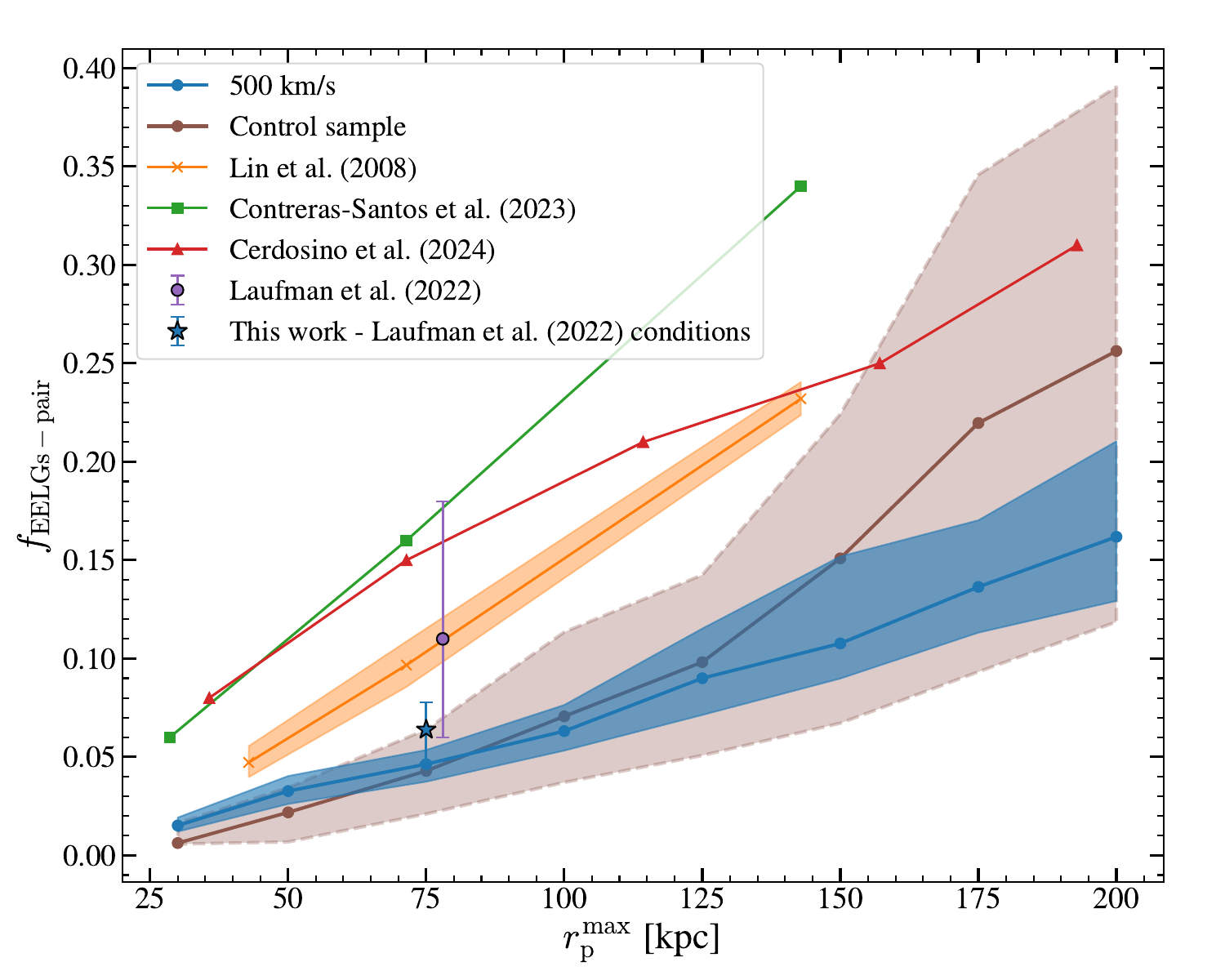}
    \caption{Same as Fig.~\ref{fig:pf_rp}, including the pair fraction obtained for the control sample (brown line and shaded area).}
    \label{fig:pf_control}
\end{figure}

To assess whether the pair fraction derived for the EELGs differs from that of other galaxies of similar stellar mass and redshift, we constructed a control sample from the J-PAS catalogue of 1.1\,M galaxies with photometric redshifts (Vázquez-Ramió et al., in prep.) and stellar masses (González-Delgado et al., in prep.) used in Sect.~\ref{sec:mass_comp}, excluding the EELGs. We divided the stellar mass--photometric redshift space into a grid covering the range spanned by the EELGs and, in each cell, randomly selected as many galaxies from the parent catalogue as EELGs in that cell. When a cell did not contain enough galaxies, all available galaxies were selected. This resulted in a control sample of 499 galaxies.

As shown in Fig.~\ref{fig:mass_completeness}, the EELGs occupy the lowest stellar-mass regime at each redshift, where the parent catalogue contains few galaxies. As a consequence, although the redshift distribution of the control sample closely reproduces that of the EELGs, the stellar-mass distribution is dominated by the more massive galaxies (Fig.~\ref{fig:control_distributions}), so that the control sample is only representative of the most massive part of the EELG population.

After applying the same selection criteria and corrections as for the EELGs, the final control sample comprises 269 galaxies. Applying the same PPF analysis, we obtain $f_\mathrm{pair}^\mathrm{control} = 0.26^{+0.13}_{-0.12}$ for $r_p^\mathrm{max}=200\,\mathrm{kpc}$ and $\Delta v_\mathrm{max}=500\,\mathrm{km\,s^{-1}}$ (Fig.~\ref{fig:pf_control}). This value is slightly higher than that obtained for the EELGs, but both are consistent within the uncertainties, which are considerably larger for the control sample owing to its smaller size and the broader photometric redshift PDFs of normal galaxies compared with those of the EELGs.

Given the limited size of the control sample and its bias towards higher stellar masses, this comparison should be regarded as a qualitative consistency check rather than a robust statistical test.

\end{appendix}
\end{document}